\documentclass[aps,prb,groupedaddress,twocolumn,superscriptaddress]{revtex4-1}
\usepackage{enumitem}
\usepackage{graphicx} 
\usepackage{bpchem}
\usepackage{dcolumn}
\usepackage{bm}
\usepackage{amsmath,amsthm,amssymb}
\usepackage{float}
\usepackage{color}
\usepackage{hyperref}
\usepackage{multirow}
\usepackage{array}

\newcommand{\cuse}{Cu$_2$OSeO$_3$}

\begin{document}
\title{Magnetic relaxation phenomena in Cu$_2$OSeO$_3$ and phase diagram}
\author{F. Qian}
\affiliation{Faculty of Applied Sciences, Delft University of Technology, Mekelweg 15, 2629 JB Delft, The Netherlands}
\author{H. Wilhelm}
\affiliation{Diamond Light Source Ltd., Chilton, Didcot, Oxfordshire, OX11 0DE, United Kingdom}
\author{A. Aqeel}
\author{T.T.M. Palstra}
\affiliation{Zernike Institute for Advanced Materials, University of Groningen, Nijenborgh 4, 9747 AG Groningen, The Netherlands}
\author{A.J.E. Lefering}
\affiliation{Faculty of Applied Sciences, Delft University of Technology, Mekelweg 15, 2629 JB Delft, The Netherlands}
\author{E. H. Br\"uck}
\affiliation{Faculty of Applied Sciences, Delft University of Technology, Mekelweg 15, 2629 JB Delft, The Netherlands}
\author{C. Pappas}
\affiliation{Faculty of Applied Sciences, Delft University of Technology, Mekelweg 15, 2629 JB Delft, The Netherlands}


\begin{abstract}
\noindent
We present an investigation of the magnetic field-temperature phase diagram of \cuse~based on DC magnetisation and AC susceptibility measurements covering a broad frequency range of four orders of magnitude, from very low frequencies reaching 0.1 Hz up to 1 kHz. The experiments were performed in the vicinity of $T_C=58.2$~K and around the skyrmion lattice A-phase. At the borders between the different phases the characteristic relaxation times reach several milliseconds and the relaxation is non-exponential. Consequently the borders between the different phases depend on the specific criteria and frequency used and an unambiguous determination is not possible. 
\end{abstract}


\maketitle
%
%

\section{Introduction}

In noncentrosymmetric magnetic materials Dzyaloshinskii-Moriya (DM) interactions \cite{Dzyaloshinsky:1958vq,Moriya:1960uf} can stabilize 2D and 3D modulations with a fixed sense of rotation of the magnetization vector. These \textit{chiral skyrmions}\cite{bogdanov1989, bogdanov1994,Rossler:2011fs,nagaosa2013} in form of axisymmetric strings have been found both in real and reciprocal space in a number of cubic helimagnets with B20 structure, such as MnSi\cite{Muhlbauer:2009bc, Tonomura:2012ep}, FeGe{\cite{Yu:2010hr, Wilhelm:2011jva, Moskvin:2013kf}, Fe$_{1-x}$Co$_x$Si\cite{Yu:2010iu, munzer2010,Park:2014jc} in the so called A-phase. This is a closed pocket of the magnetic field ($B$) - temperature ($T$) phase diagram. The recent discovery of similar behaviour in the insulator and multiferroic \cuse\cite{Seki:2012ie,Seki:2012ch,Adams12} has attracted attention also because in this system it is possible to manipulate skyrmions by external electric fields\cite{White12,White14}.

\cuse~crystallizes in the non-centrosymmetric space group P2$_1$3, the same as for the B20 compounds, but with two different Cu$^{2+}$ ion sites\cite{Bos08,Belesi:2010vz}, as shown in the inset of Fig.~\ref{fig:crystalstructure}. The balance between the ferromagnetic exchange and the DM interactions leads to a long-period helical order with a pitch of $\sim70$~nm \cite{Seki:2012ie,Adams12}. A weak anisotropy fixes the helices along the $\langle100\rangle$ crystallographic directions below the ordering temperature $T_C$, which is close to 58~K. A weak external magnetic field $B$ may overcome the anisotropy, unpin the helices from the lattice and orient them along its direction leading to the conical phase if $B>B_{C1}$. Higher magnetic fields stabilize the A-phase pocket close to $T_C$ and even higher magnetic fields are needed to overcome the DM interaction and the helical correlations inducing the field polarised phase, which sets-in for $B>B_{C2}$\cite{Seki:2012ie,Adams12}.
These features are summarised in Fig.~\ref{fig:crystalstructure}, which schematically illustrates the spin arrangements of the various phases.

The phase diagram of \cuse~has been investigated close to $T_C$ by neutron scattering\cite{Seki:2012ie,Seki:2012ch, Adams12} and AC susceptibility for frequencies from 2.5 Hz to 1 kHz\cite{Levatic14}. The DC magnetisation and AC susceptibility measurements presented in the following complement these previous studies. Our experiments span a very broad frequency range, from 0.1~Hz to 1~kHz, which extends the previous study\cite{Levatic14} towards the low frequencies by more than one order of magnitude. The analysis of the results as a function of the magnetic field, instead of the temperature as in the previous study, provides a quantitative approach to the phase diagram. The dynamics at the transitions between the helical, conical and A-phases, at $B_{C1}$, $B_{A1}$ and $B_{A2}$, involve a broad distribution of relaxation times with characteristic times reaching several milliseconds. Additional relaxation processes have also been found at very low frequencies and around $B_{C1}$. The borders between the different phases are discussed and we conclude that these are not sharp but their exact location depends on the specific criteria and the frequency used. Remarkably, no relaxation is found at the high and the low temperature boundaries of the A-phase. 

\begin{figure}
\includegraphics[width= 0.4\textwidth]{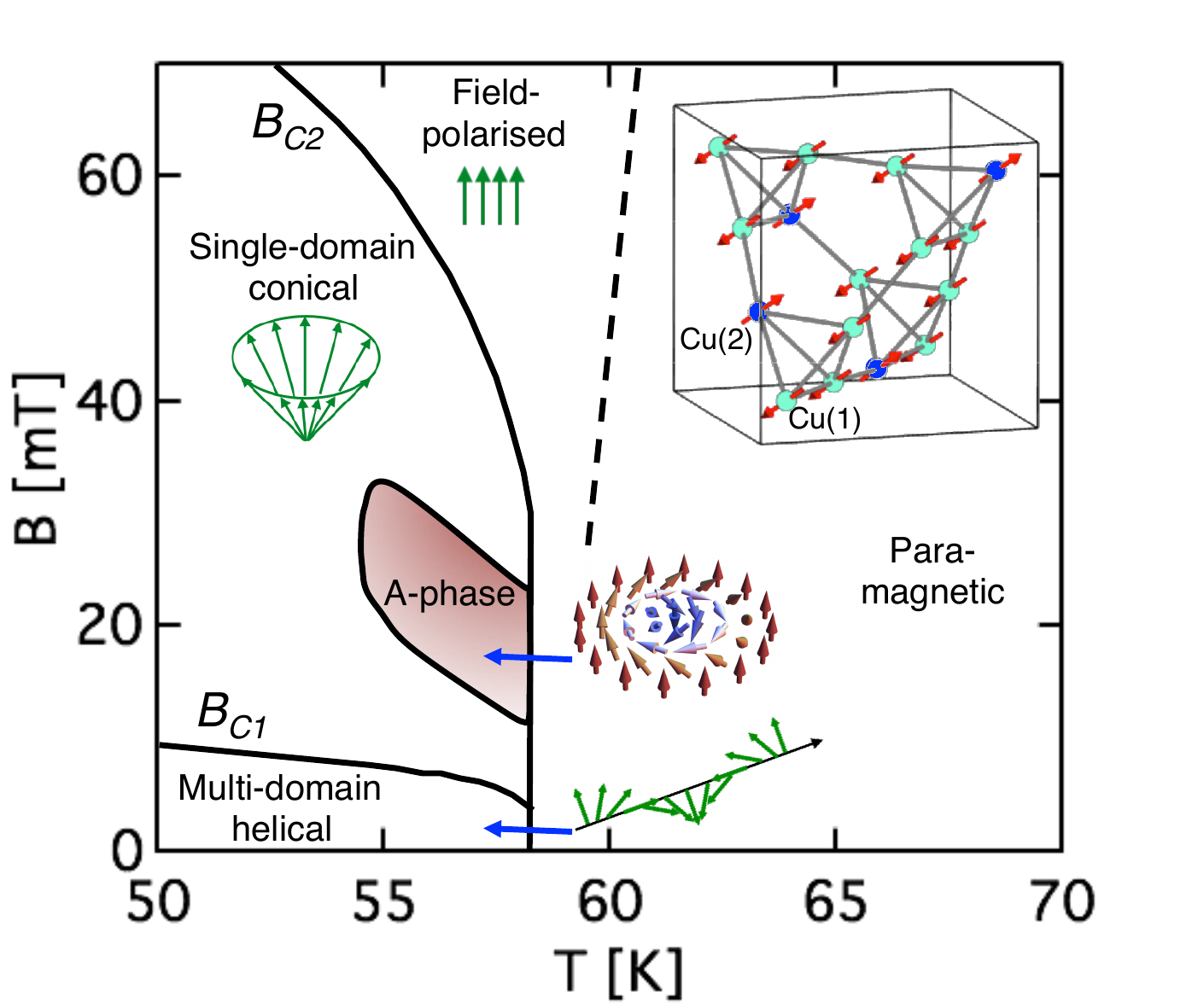}
\caption{
Schematic representation of the phase diagram of \cuse~illustrating the spin arrangements of the various phases. The inset shows the crystal structure with the two different Cu$^{2+}$ ion sites\cite{Bos08}.}
\label{fig:crystalstructure}
\end{figure}

The manuscript will present and discuss the experimental results at separate sections: experimental details, magnetization, AC susceptibility at a frequency of 0.8 Hz, frequency dependence, Cole-Cole analysis and finally the $B-T$ phase diagram. 

\section{Experimental Details}
\label{Experimental Details}
A high-quality single crystal of \cuse~was grown by the chemical vapour transport method\cite{Miller:2010fu} and its structure was checked by X-ray diffraction. The sample with almost cubic shape was oriented with the $(001)$ axis vertical within $\pm5^\circ$. The magnetization $M$ and the real and imaginary components of the AC-susceptibility, $\chi'$ and $\chi''$ respectively, were measured with a MPMS-XL Quantum design SQUID magnetometer using the extraction method. A DC magnetic field $B$ was applied along the vertical direction and parallel to the drive AC field, $B_{AC}$, with $0< B_{AC} \leq 0.4$ mT, within the frequency range $0.1\leq f \leq 1000$~Hz. Frequency scans were performed with logarithmic steps for each $B$ and after having checked that the AC susceptibility was independent of $B_{AC}$, the measurements were done for $B_{AC}$ = 0.4~mT. 
For the measurements two specific experimental protocols have been used:
\begin{itemize} [noitemsep]
\item  FC temperature scans: the sample was brought to 70~K, a magnetic field was applied and the signal was recorded with decreasing stepwise the temperature. At each temperature the sample was brought to thermal equilibrium before measurement. 
\item  ZFC magnetic field scans: the sample was brought to the temperature of interest under zero field (more specifically the residual field of the magnetometer which was less than 1~mT). Once thermal equilibrium was reached the measurements were performed by increasing stepwise the magnetic field.
\end{itemize}
%
%
\section{Magnetization}
\label{magnetisation}
Figure~\ref{fig:MvsB_plots} displays the ZFC magnetization $M$ and the susceptibility $\Delta M/\Delta B$, derived by numerical differentiation of the $M$ vs $B$ curves. The same set of data has been plotted either versus the magnetic field for selected temperatures (a and c) or versus the temperature for selected magnetic fields (b and d) in order to better emphasise the occurring phase transitions. The magnetisation curves in panels (a) and (b) highlight the transition from the paramagnetic to the helical phase around 58~K, seen both in the strong non-linearities of $M$ versus $B$ for $56 \text{~K} \lesssim T\lesssim 59\text{~K}$ (a) and the onset of a plateau at the M versus T plots (b). The $M$ vs $B$ plots give only a coarse overview and the effects related with the onset of the $A$-phase, are only brought out by the $\Delta M/\Delta B$ curves, which show clear dips at $56 \text{~K} \lesssim T\lesssim 58\text{~K }$ and $15 \text{~mT} \lesssim B\lesssim 30\text{~mT}$. 
Thus the relevant features are seen on $\Delta M/\Delta B$ and show up with much higher accuracy on the AC susceptibility which will be discussed in the following sections. 
%
\begin{figure}
\includegraphics[width= 0.4\textwidth]{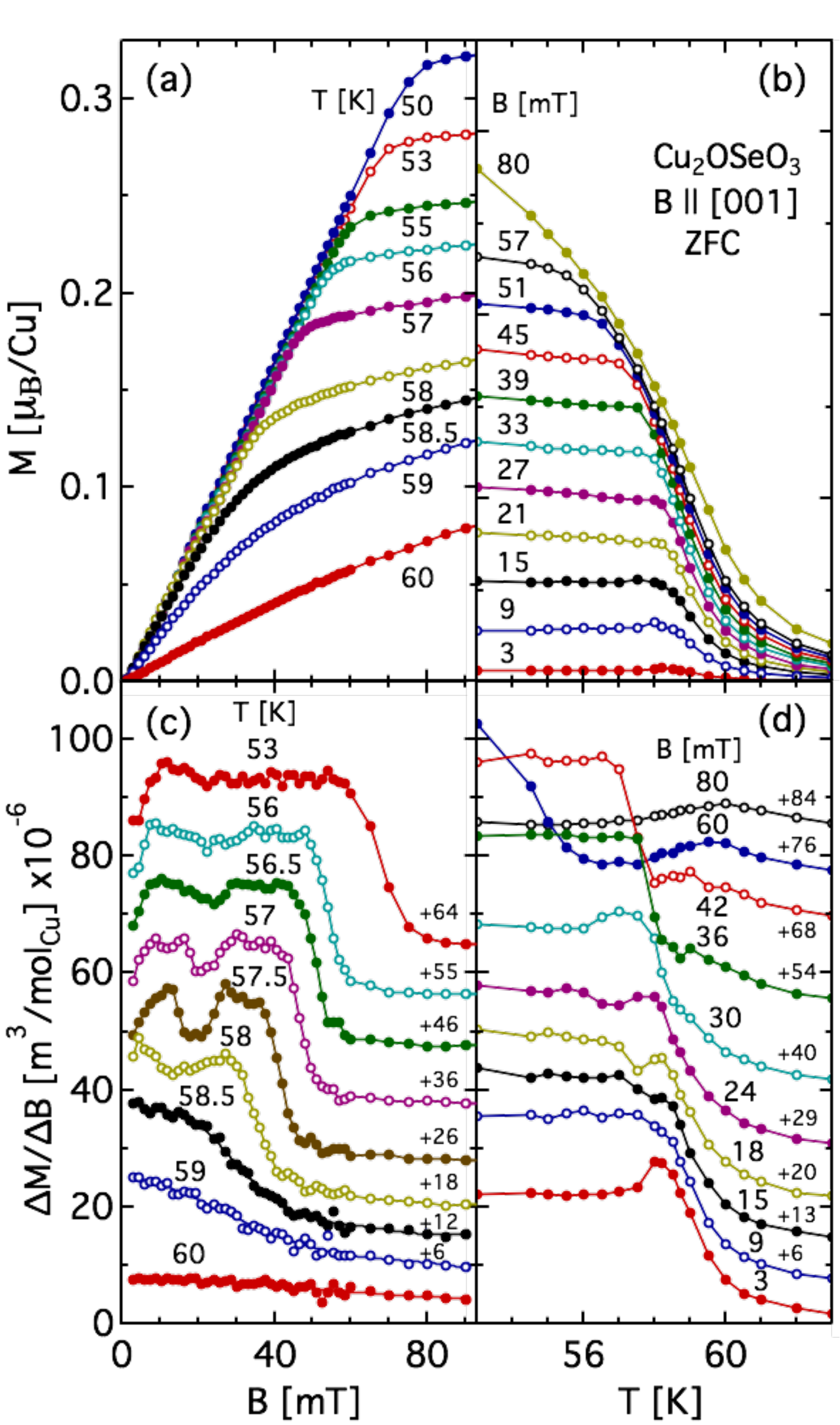}
\caption{
ZFC magnetization of \cuse~as a function of temperature and magnetic field. The same set of data has been plotted versus the magnetic field for selected temperatures (a) and versus the temperature for selected magnetic fields (b). The susceptibility $\Delta M/\Delta B$ deduced by numerical differentiation of the $M$ vs $B$ curves is shown as a function of the magnetic field in (c) and of temperature in (d). For the sake of clarity the curves in (c) and (d) have been shifted vertically with respect to the baseline by the values indicated.}
\label{fig:MvsB_plots}
\end{figure}
\section{ AC susceptibility at 0.8~$\text{Hz}$}
\label{susceptibility at 0.8}
\begin{figure}
\includegraphics[width= 0.35 \textwidth]{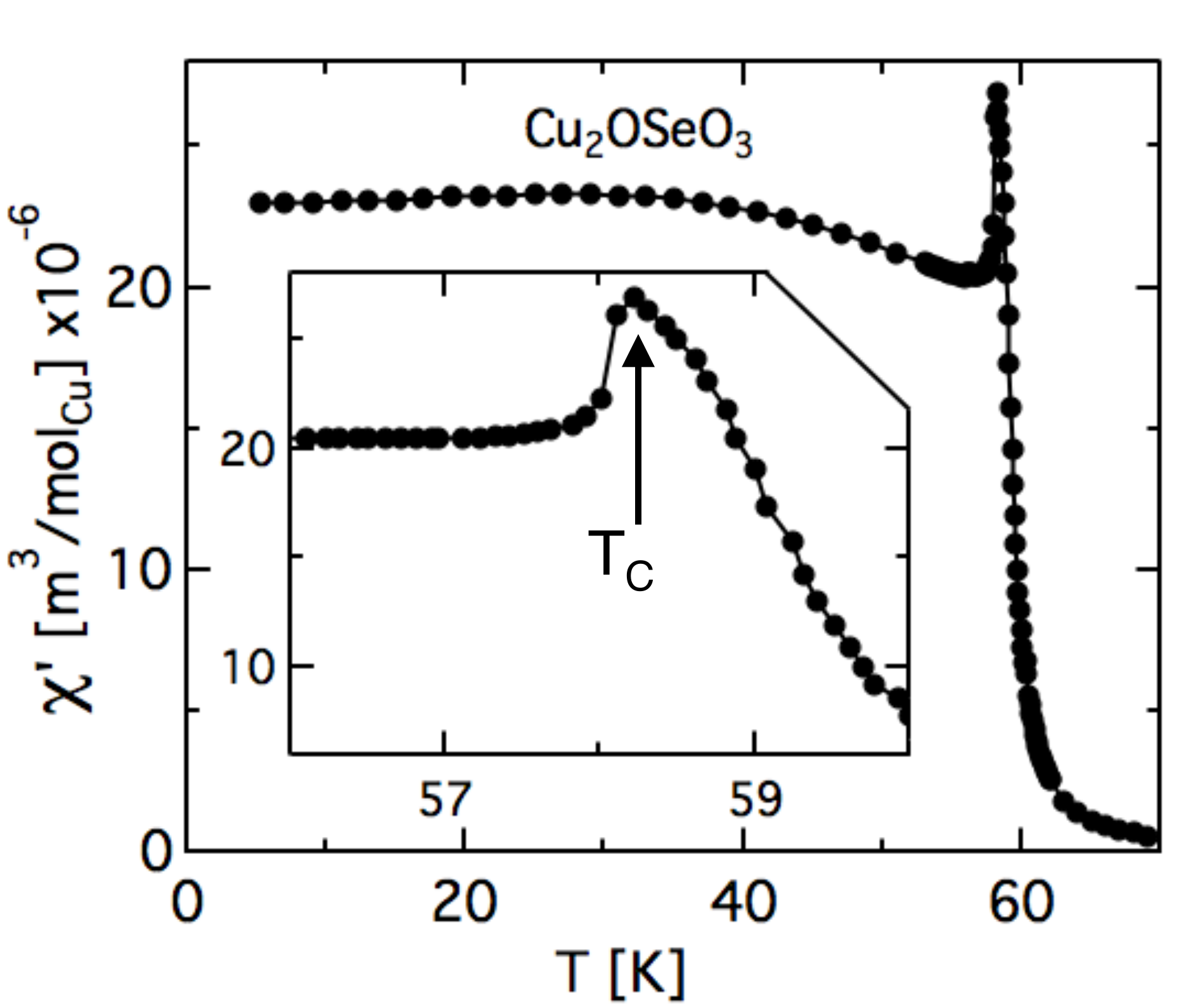}
\caption{ 
Temperature dependence of $\chi'$  of \cuse~for $B=0$~mT and $f=0.8$~Hz. The inset shows a close-up around the peak, which marks the onset of the helical order and reveals an asymmetric shape with a shoulder above $T_C=58.20 \pm 0.05 $~K.}
\label{fig:chiVsTzeroH}
\end{figure}
%
Figure~\ref{fig:chiVsTzeroH}(a) shows the temperature dependence of $\chi'$ measured at $B=0$~mT and at a frequency of 0.8~Hz. For $B=0$~mT and all frequencies used in this work $\chi'$ was frequency independent and $\chi''=0$. The peak at $T_C=58.20 \pm 0.05$~K marks the transition to the helical state and the closer inspection shown in the inset reveals an asymmetric cusp-like shape with a shoulder at  59$\pm$0.01~K, as reported before in \cuse~\cite{Levatic14,Zivkovi12}. Similarly to MnSi\cite{2007PhRvB..76e2405S} the shoulder marks the onset of the precursor phase \cite{2016arXiv160606922S}, where helical correlations and fluctuations become predominant. 

In the helical phase $\chi'$ varies non-monotonically showing a minimum at $\sim$ 57~K and a broad maximum at about 30~K before levelling-off to $\sim23\cdot10^{-6}\ \text{m}^3\text{/mol}_{\mathrm{Cu}}$ below 10~K. This value is comparable to the one reported for a polycrystalline sample\cite{Bos08}, where however a smoother temperature dependence with no clear peak at $T_C$ has been found. On the other hand, the overall shape of $\chi'$ in Fig.~\ref{fig:chiVsTzeroH} is similar to the one found for another single crystal along  $\langle$111$\rangle$, where slightly lower absolute values have been reported possibly due to the different crystal orientations\cite{Zivkovi14}.
\begin{figure}
\begin{center}
\vspace*{0mm}
\hspace*{0mm}
\includegraphics[width= 0.45 \textwidth]{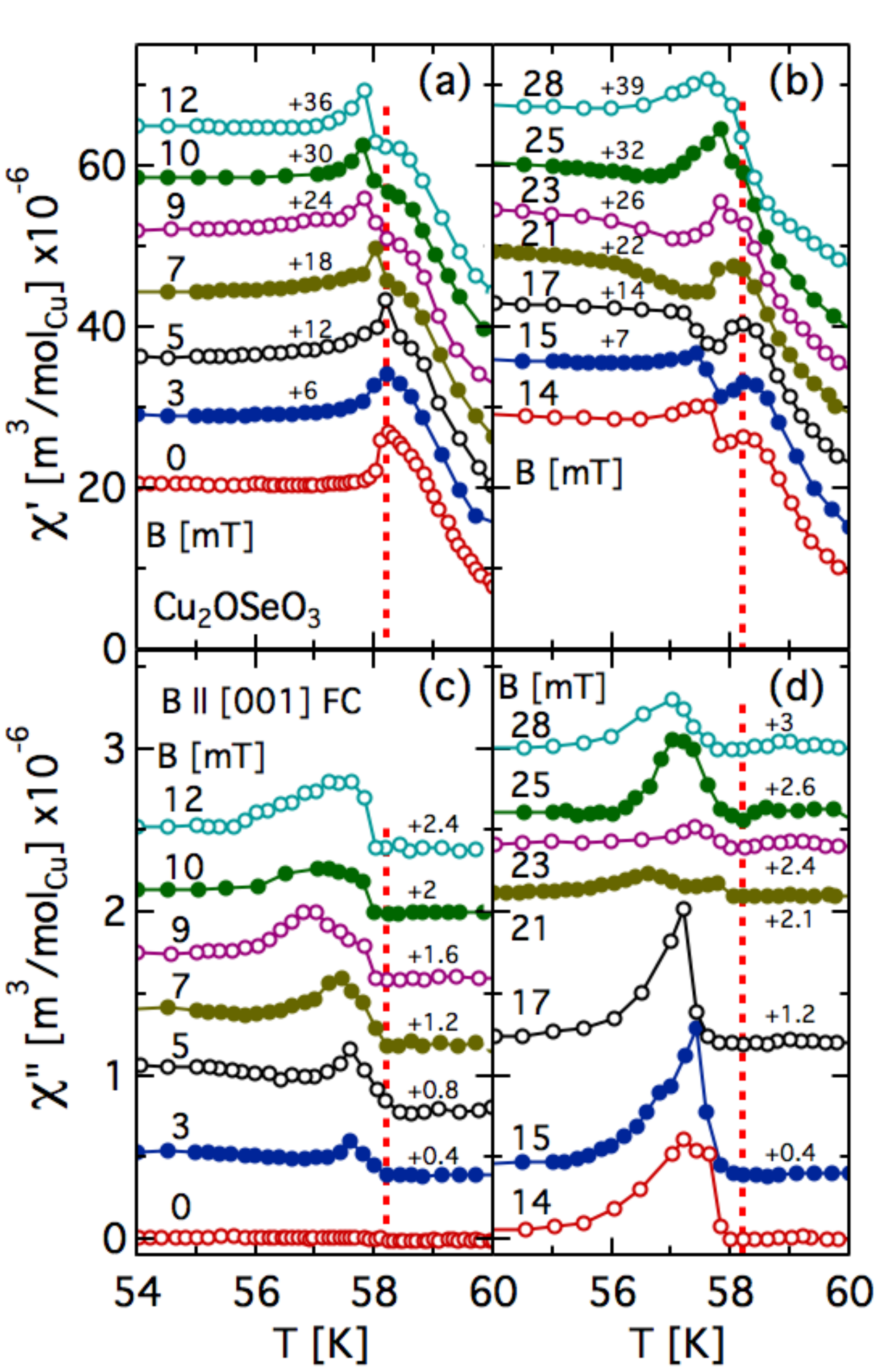}
\caption{
Temperature dependence of FC $\chi'$ (a and b) and $\chi''$ (c and d), for different magnetic fields $B$ and for $f=0.8$~Hz. The vertical dashed lines indicate $T_C$ for $B=0$. For the sake of clarity the curves have been shifted vertically with respected to the base line by the numbers given next to each of them.}
\label{fig:chiVsT} 
\end{center}
\end{figure}
\begin{figure}[t]
\begin{center}
\vspace*{0mm}
\hspace*{-5mm}
\includegraphics[width= 0.35 \textwidth]{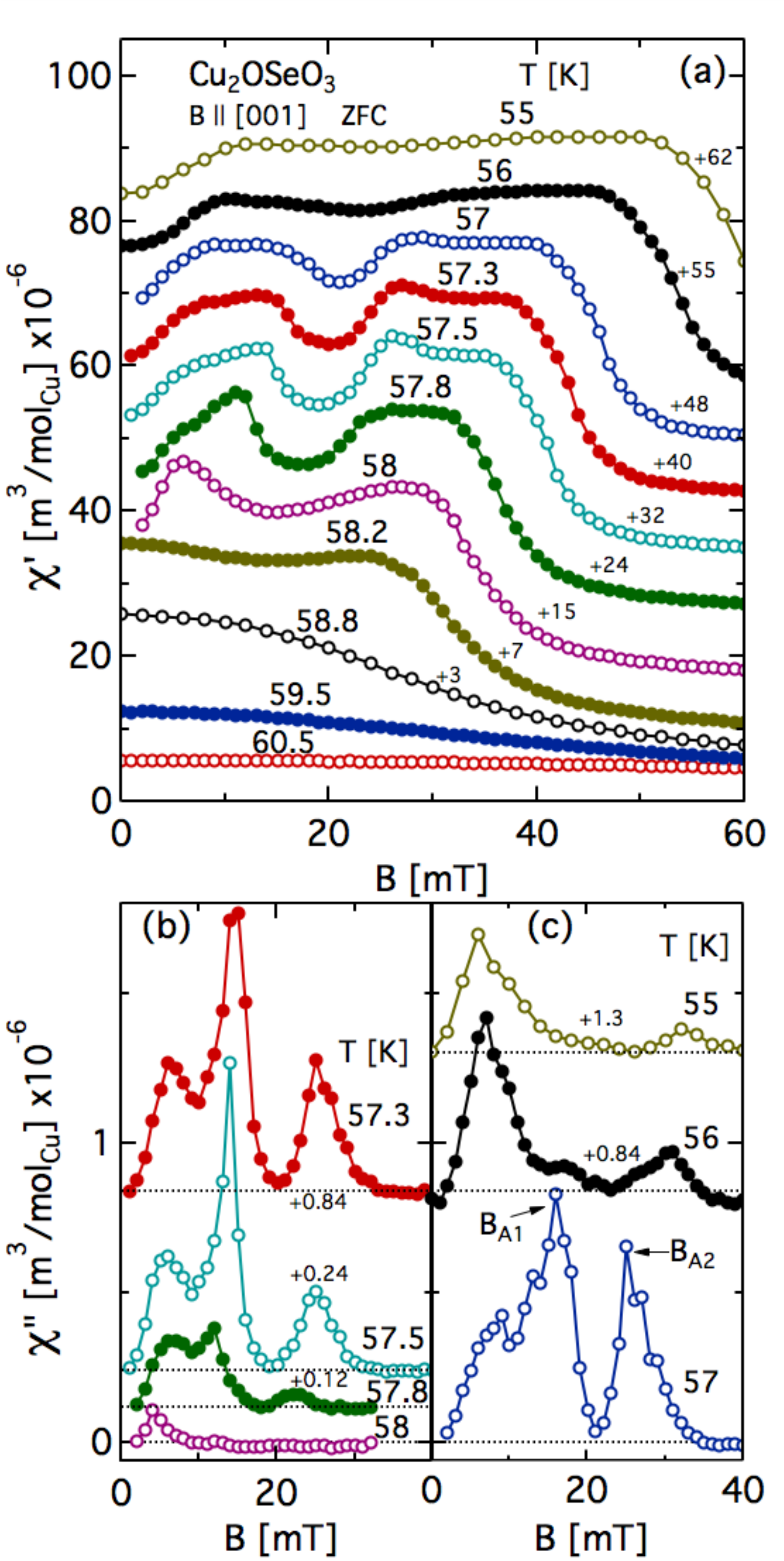}
\caption{ 
Magnetic field dependence of ZFC $\chi'$ (a) and $\chi''$ (b and c) of \cuse~for some selected temperatures and for $f=0.8$~Hz. For the sake of clarity some curves have been vertically shifted with respect to the base line by the values indicated next to each of them.}
\label{fig:chiVsB}
\end{center}
\end{figure}

At low temperatures the magnetic response depends on the magnetic history of the sample. It has indeed been found that below 50~K FC and ZFC measurements give different results\cite{Bos08,Zivkovi12}. For this reason we distinguished between FC or ZFC following the specific procedures described above. The results did not show any influence of the magnetic history on the susceptibility. Despite of that, for the sake of clarity, we will specify in the following the protocol used for each set of data. 

Figure~\ref{fig:chiVsT} shows the FC susceptibility, $\chi'$ (a and b) and $\chi''$ (c and d), in the vicinity of $T_C$ for selected values of $B$ and reveals the strong influence of the magnetic field: the peak of $\chi'$, which has been associated with $T_C$ at $B=0$, shifts to lower temperatures whereas the high-temperature shoulder becomes more noticeable. At $12\text{~mT}\lesssim B\lesssim 23\text{~mT}$ the shape of $\chi'$ changes dramatically: it shows two maxima separated by a minimum characteristic for the $A$-phase. By further increasing the magnetic field only one cusp remains and the shape becomes again similar to that of low fields. Much higher magnetic fields smoothen the cusp and $\chi'$ becomes almost temperature independent below $\sim58$~K. 

The magnetic field has a dramatic influence also on $\chi''$. The weakest magnetic field (3~mT in our case) induces already a peak in $\chi''$ slightly below $T_C$, which becomes more pronounced at 5~mT and transforms into a broad maximum upon further increasing the field. Between 14 and 17~mT the maximum becomes a strong and well defined cusp. By further increasing $B$, $\chi''$ vanishes for 21~mT \textless$B$~\textless23~mT but re-appears for $B$ \textgreater 23~mT before fading away at higher fields approaching $B_{C2}$. 
 
A complementary view of the effect of temperature and magnetic field is given by Fig.~\ref{fig:chiVsB}, where the ZFC susceptibility is plotted versus the magnetic field for selected temperatures. Well above $T_C$, at 60.5 K, $\chi'$ is almost field independent and $\chi''$ is practically zero. Lowering the temperature leads to a strong increase of the low field $\chi'$. Furthermore, at $T_C$ a non-monotonic behaviour develops with a minimum at $B\sim 14$~mT, which is characteristic for the A-phase. The minimum is most pronounced for $58 \text{ K} \gtrsim T \gtrsim 57 \text{ K}$ and persists down to 55~K. 

As already mentioned $\chi''$ is zero for $T\geq T_C$ and for all frequencies used in this study. A peak first appears at 58~K centered at $B=4 \text{ mT}$ and remains significant upon decreasing the temperature. At $57.8$~K two additional maxima appear at the borders of the $A$-phase, which evolve to well defined peaks at 57~K and fade out at lower temperatures. 
%
\begin{figure}
\begin{center}
\vspace*{0mm}
\hspace*{-5mm}
\includegraphics[width= 0.35 \textwidth]{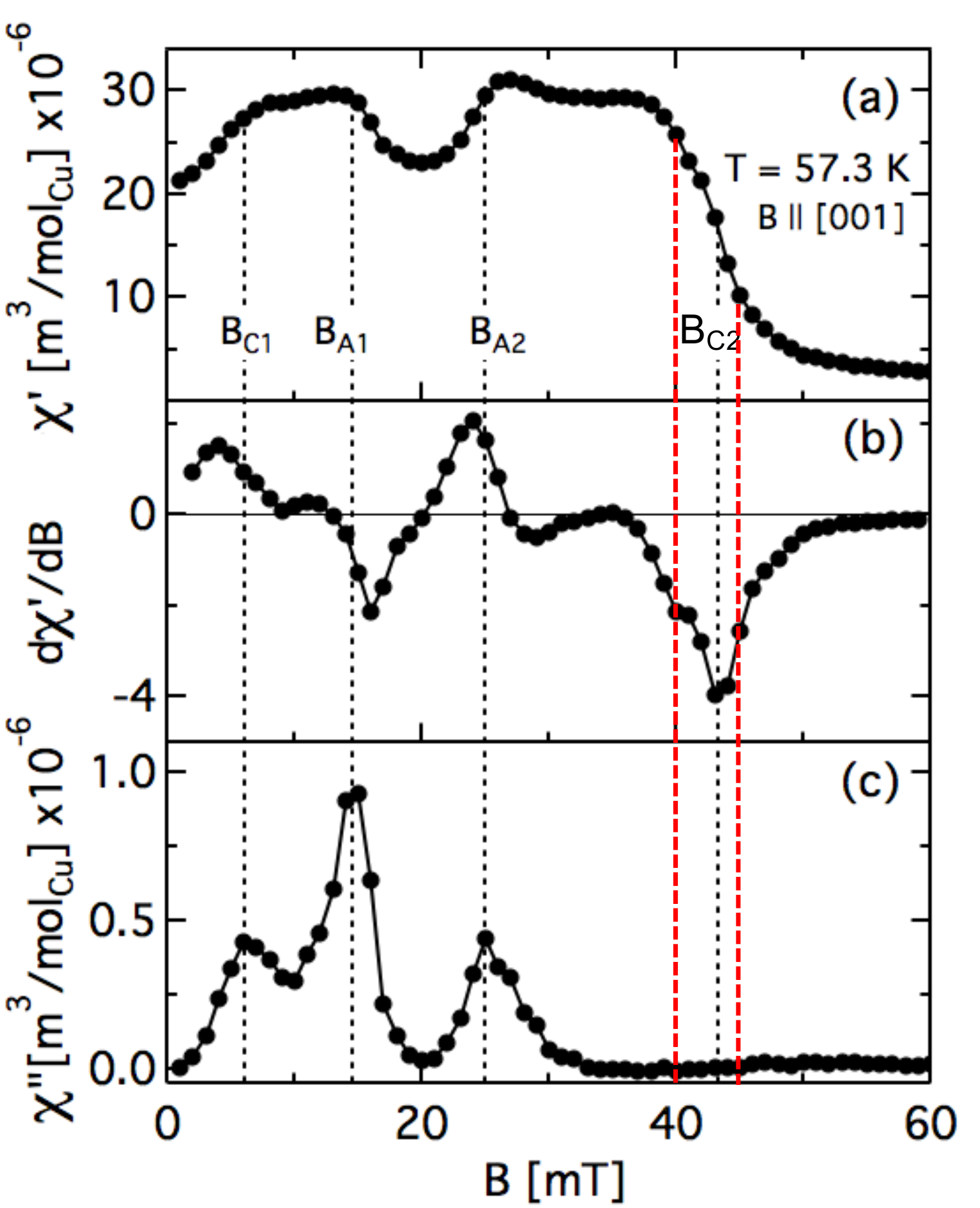}
\caption{
Magnetic field dependence of ZFC $\chi'$, its derivative $d\chi'/dB$ and the corresponding $\chi''$ of \cuse~at $T=57.3$~K for $B\parallel [001]$ and $f=0.8$~Hz. The maxima of $\chi''$ define the lower critical field $B_{C1}$ = 6~mT as well as the lower and upper boundaries of the $A$-phase, $B_{A1}$ = 14.5~mT and $B_{A2}$ = 25~mT, respectively. The upper critical field ($B_{C2}$ = 43 mT) was determined from the inflection points of $\chi'$ and the extrema of $d\chi'/dB$. The red dotted lines correspond to the inflection points of  $d\chi'/dB$ at the two sides of $B_{C2}$ and thus to the lower and upper boundaries of this transition. The unit of $d\chi'/dB$ is m$^3$/(mol$_{\text{Cu}}\cdot\text{mT})\times10^{-6}$.}
\label{fig:definitionBoundaries}
\end{center}
\end{figure}

Both $\chi'$ and $\chi''$ bear the signature of the series of field-induced transitions characteristic of \cuse\cite{Seki12, Adams12, Levatic14} and helimagnets of B20 types such as MnSi{\cite{Muhlbauer:2009bc, Bauer:2012cw} or FeGe\cite{Wilhelm:2011jva, Moskvin:2013kf}. Below $T_C$ the application of a magnetic field induces an initial increase of $\chi'$ related to the transition from the helical to the conical phases at the lower critical field $B_{C1}$. By further increasing the magnetic field at sufficiently low temperatures, for \cuse~at $T<$ 55 K, $\chi'$ remains constant until the upper critical field $B_{C2}$, where it decreases rapidly indicating the transition to the field polarised state. Besides this generic scheme, very close to $T_C$ additional features appear both in $\chi'$ and $\chi''$ in relation with the boundaries of the A-phase. 

The extraction of an exact phase diagram however, with precise values for the critical fields is not an easy task as the result very much depends on the specific criteria used. This is illustrated by Fig.~\ref{fig:definitionBoundaries}, that displays the magnetic field dependence of the ZFC $\chi'$, its first derivative $d\chi'/dB$ and the corresponding $\chi''$ at $T=57.3$~K, a characteristic temperature where all features are present. One may indeed choose either the inflection points of $\chi'$ or the maxima of $\chi''$ to define the phase boundaries. Both choices are valid and would lead to magnetic fields with similar temperature dependence. We chose the peaks of $\chi''$ at $f=0.8$~Hz to define $B_{C1}$ and the boundaries of the $A$-Phase, $B_{A1}$ and $B_{A2}$ respectively. On the other hand $\chi''$ is zero at high fields and for this reason $B_{C2}$ was defined from the inflection point of $\chi'$. The lower and upper boundaries of this transition can be estimated from the extrema of the second derivative $d^2\chi'/dB^2$ (thus the inflection points of the first derivative $d\chi'/dB$), which are given by the red dotted lines in Fig. \ref{fig:definitionBoundaries}(b) and will discussed at the phase diagram section below.

A clear overview of the phase diagram and the transitions between the helical, conical, A- and field polarized phases respectively is only provided by the  analysis as a function of the magnetic field presented above. If only the temperature dependence of the susceptibility is considered, as it was the case in the previous study\cite{Levatic14}, the different contributions merge into broad features as illustrated by  Fig.~\ref{fig:chiVsT}(c), in particular between 9 and 12 mT, and thus no clear indications for the phase boundaries can be obtained. 

\section{Frequency Dependence of $\chi'$ and $\chi''$}
\label{sec:resultsFreqdependence}
\begin{figure}
\begin{center}
\includegraphics[width= 0.35 \textwidth]{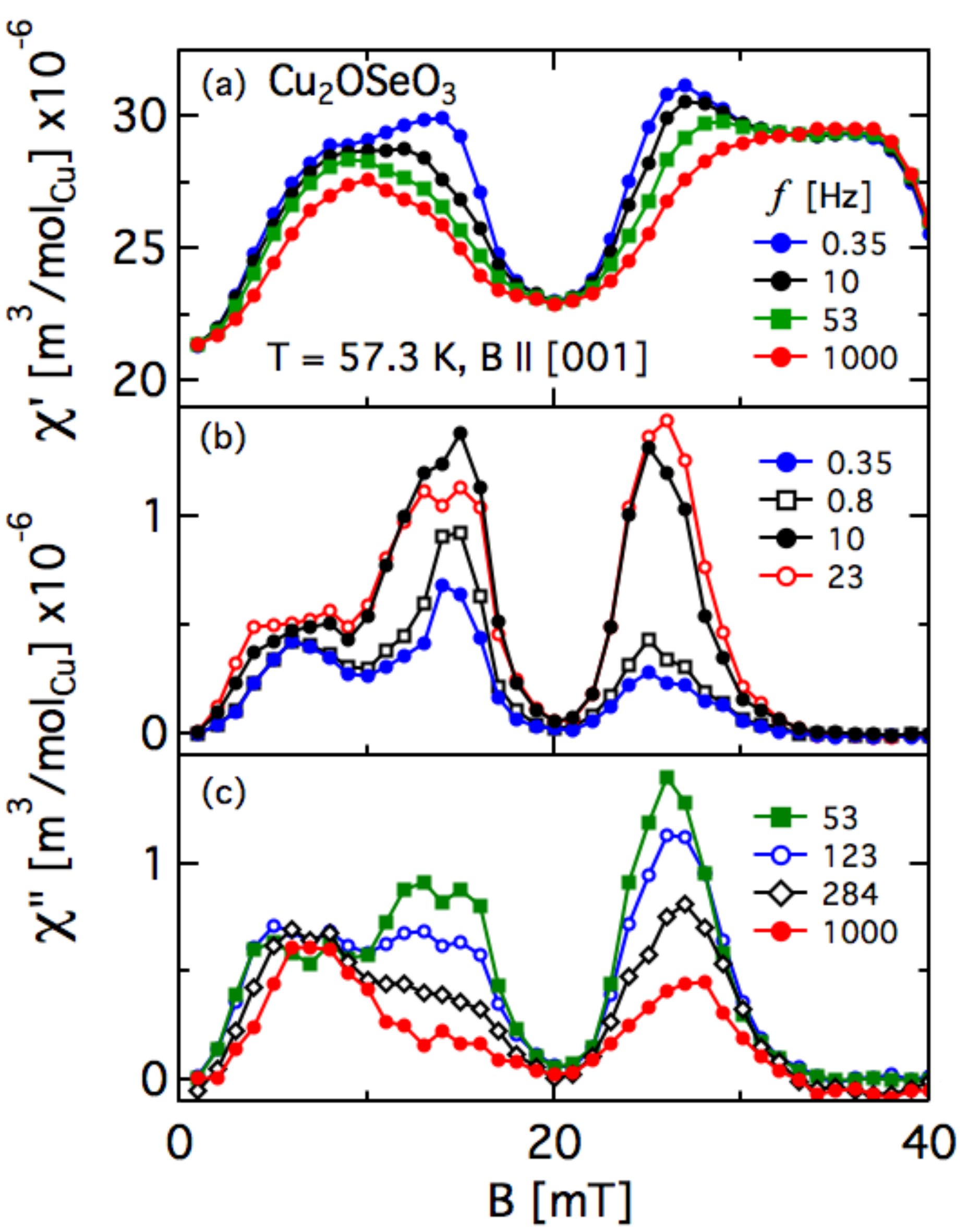}
\caption{
Magnetic field dependence of ZFC $\chi'$ (a) and $\chi''$ (b and c) of \cuse~ at $T=57.3$~K and for the frequencies indicated.}
\label{fig:chiVsBvariousFreq}
\end{center}
\end{figure}
%
\begin{figure}
\begin{center}
\includegraphics[width= 0.35 \textwidth]{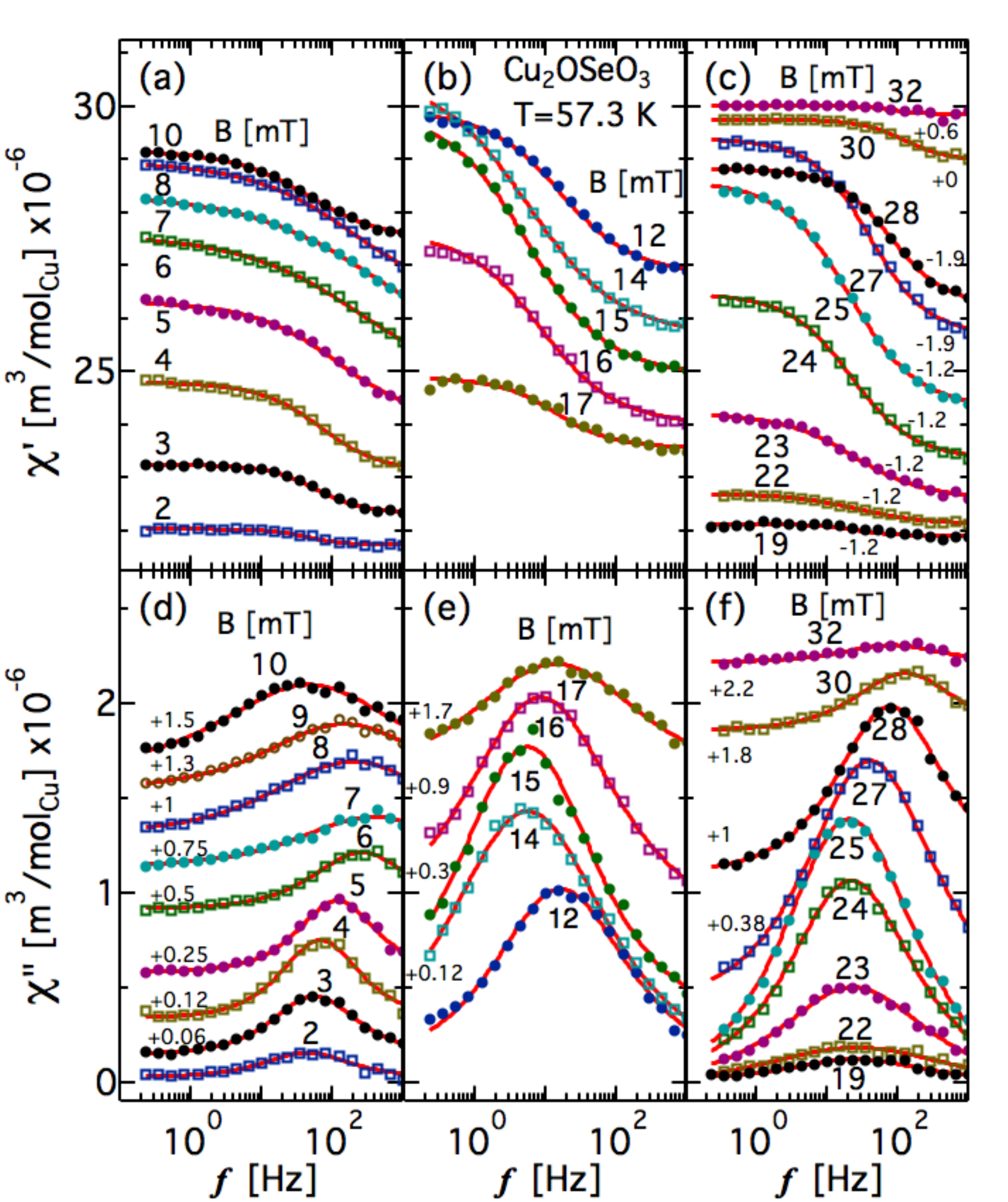}
\caption{
Frequency dependence of ZFC $\chi'$ and $\chi''$ of \cuse~at $T=57.3$~K for magnetic fields around $B_{C1}$ (a and d), $B_{A1}$ (b and e) and $B_{A2}$ (c and f) respectively. The lines represent fits of eqs.~(\ref{eq:Chi'Distribution}) and (\ref{eq:Chi''Distribution}). For the sake of clarity some data sets in panel (c-f) have been vertically shifted with respect to the baseline as indicated.}
\label{fig:chiVsFreqVariousB}
\end{center}
\end{figure}
%
The previous section discussed results at $f=0.8$~Hz. However, the susceptibility depends on the frequency of the AC drive field as highlighted by Fig.~\ref{fig:chiVsBvariousFreq}, where ZFC $\chi'$ and $\chi''$ are displayed versus the magnetic field for some selected frequencies and $T=57.3$~K. The minimum in $\chi'$, which is characteristic for the A-phase is best defined at low frequencies. With increasing frequency, the minimum broadens, accompanied by a decreasing amplitude of the hump-like edges at both sides. 
The most dramatic changes are found for $\chi''$, which at 0.35~Hz displays the three well-defined maxima related to the phase boundaries $B_{C1}$, $B_{A1}$ and $B_{A2}$. The amplitude of these maxima changes with frequency but their position remains roughly the same. 

A more detailed overview of the effect of frequency is given in Fig.~\ref{fig:chiVsFreqVariousB}, where $\chi'$ and $\chi''$ are displayed as a function of frequency for selected magnetic fields around $B_{C1}$~= 6~mT (a and d), $B_{A1}$~= 14.5~mT (b and e) and $B_{A2}$~= 25~mT (c and f) at $T$ = 57.3 K. As already mentioned and illustrated by Fig.~\ref{fig:chiVsBvariousFreq}, $\chi'$ does not depend on the frequency for $B=0$ and $\chi''=0$. However, even weak magnetic fields induce noticeable effects shown in both Fig.~\ref{fig:chiVsBvariousFreq} and \ref{fig:chiVsFreqVariousB}. The broad bell-shaped frequency dependence of $\chi''$ seen in Fig.~\ref{fig:chiVsFreqVariousB} (d-f) reflects a distribution of relaxation frequencies centred at the characteristic frequency $f_0$, which varies non-monotonically around $B_{C1}$, $B_{A1}$ and $B_{A2}$. Similar behaviour has already been reported for \cuse\cite{Levatic14} and the quantitative analysis is based on the modified Cole-Cole formalism\cite{cole_cole, Hueser86}:
%
\begin{figure}
\begin{center}
\includegraphics[width= 0.35 \textwidth]{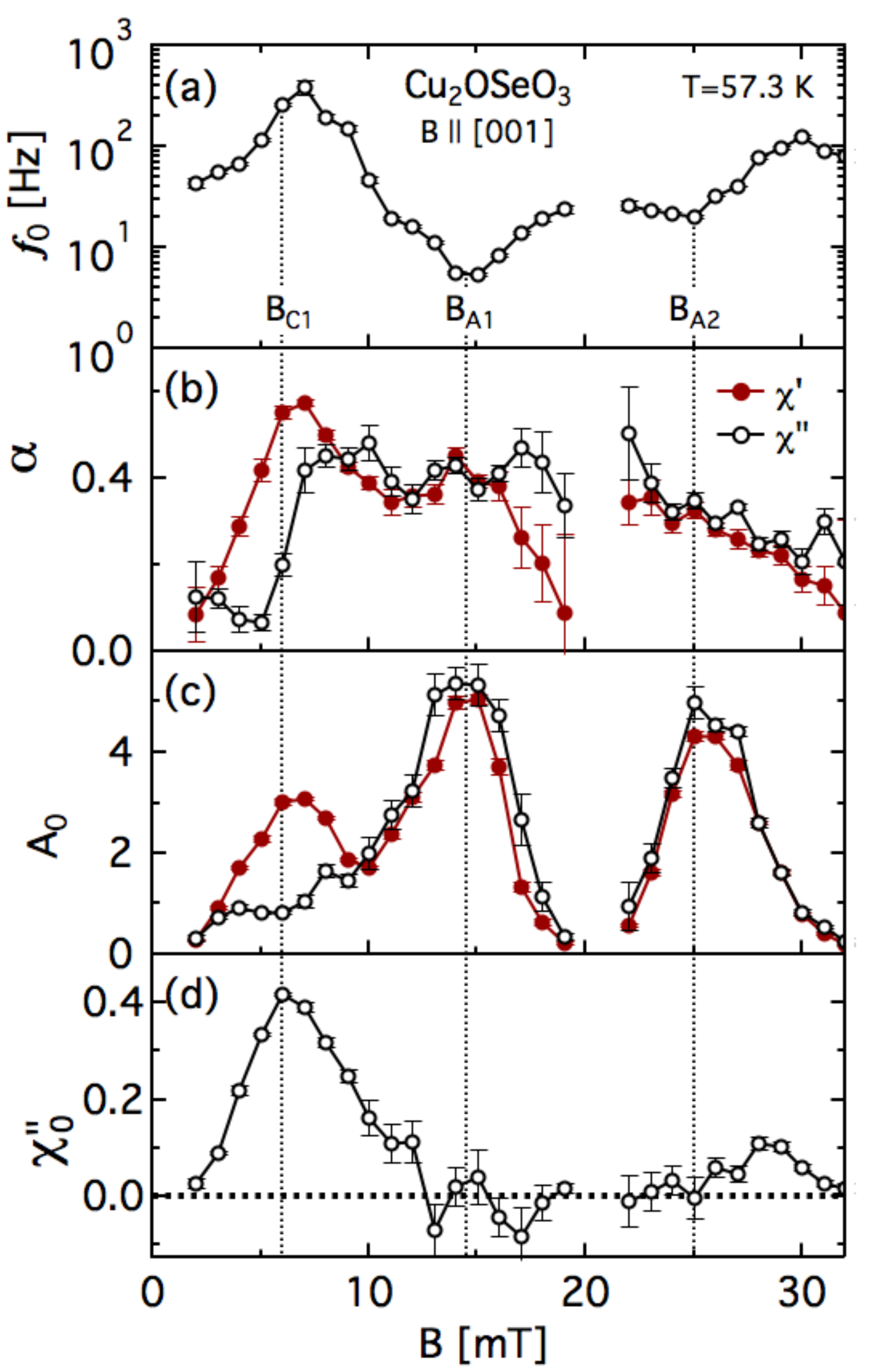}
\caption{
Magnetic field dependence of (a) the characteristic frequency $f_0$, (b) the relaxation times distribution parameter $\alpha$, and (c) $A_0=\chi(0)-\chi(\infty)$ 
as defined by eq.~\ref{eq:Chi'Distribution}. The open and closed symbols correspond to the parameters extracted from $\chi''$ and $\chi'$ respectively. The phase boundaries $B_{C1}$, $B_{A1}$, and $B_{A2}$ are indicated by the vertical dashed lines. No fits are possible in the centre of the A-phase, which explains the absence of points. Panel (d) displays a frequency independent component $\chi''_0$, which is required to fit $\chi''$ around $B_{C1}$ and accounts for the different $A_0$ derived from $\chi'$ and $\chi''$ respectively. The units of A$_0$ and $\chi''_0$ are m$^3$/mol$_{\text{Cu}}\times$10$^{-6}$.}
\label{fig:fitParametersT57p3K}
\end{center}
\end{figure}
\begin{figure}
\begin{center}
\includegraphics[width= 0.35 \textwidth]{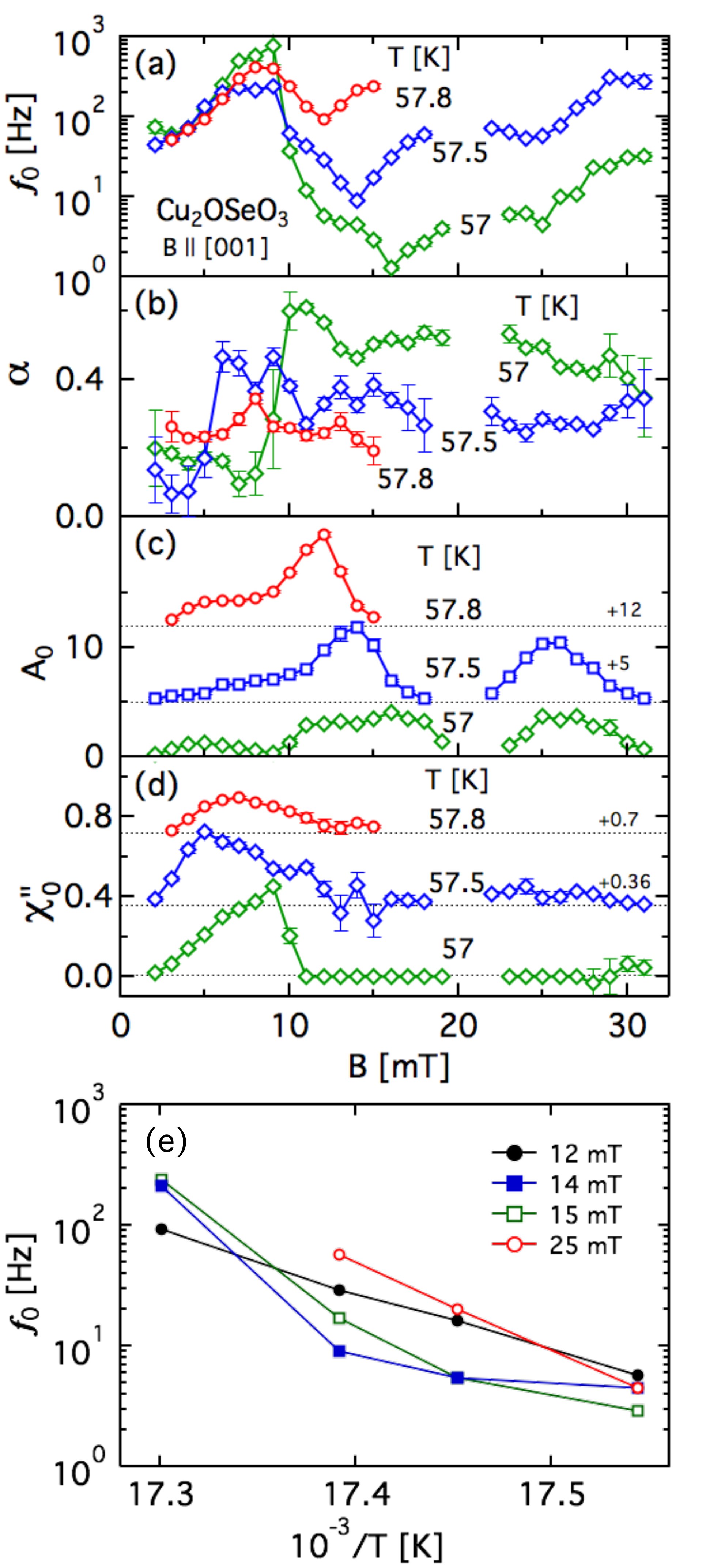}
\caption{
Magnetic field dependence of (a) the characteristic frequency $f_0$, (b) the relaxation times distribution parameter $\alpha$, (c) the amplitude $A_0=\chi(0)-\chi(\infty)$ and (d) the constant term $\chi''_0$ for the temperatures indicated, obtained from the fits of eq.~\ref{eq:Chi''Distribution} to $\chi''$ of \cuse. For the sake of clarity some $A_0$ and $\chi''_0$ curves have been shifted with respect to the baseline as indicated. No fits are possible in the centre of the A-phase and above 15 mT for 57.8 K, which explains the absence of points. The units of A$_0$ and $\chi''_0$ are m$^3$/mol$_{\text{Cu}}\times$10$^{-6}$. Panel (e) shows the temperature dependence of $f_0$ from panel (a)  for various magnetic fields. }
\label{fig:fit_Params_vs_B.pdf}
\end{center}
\end{figure}
\begin{figure}
\begin{center}
\includegraphics[width= 0.35 \textwidth]{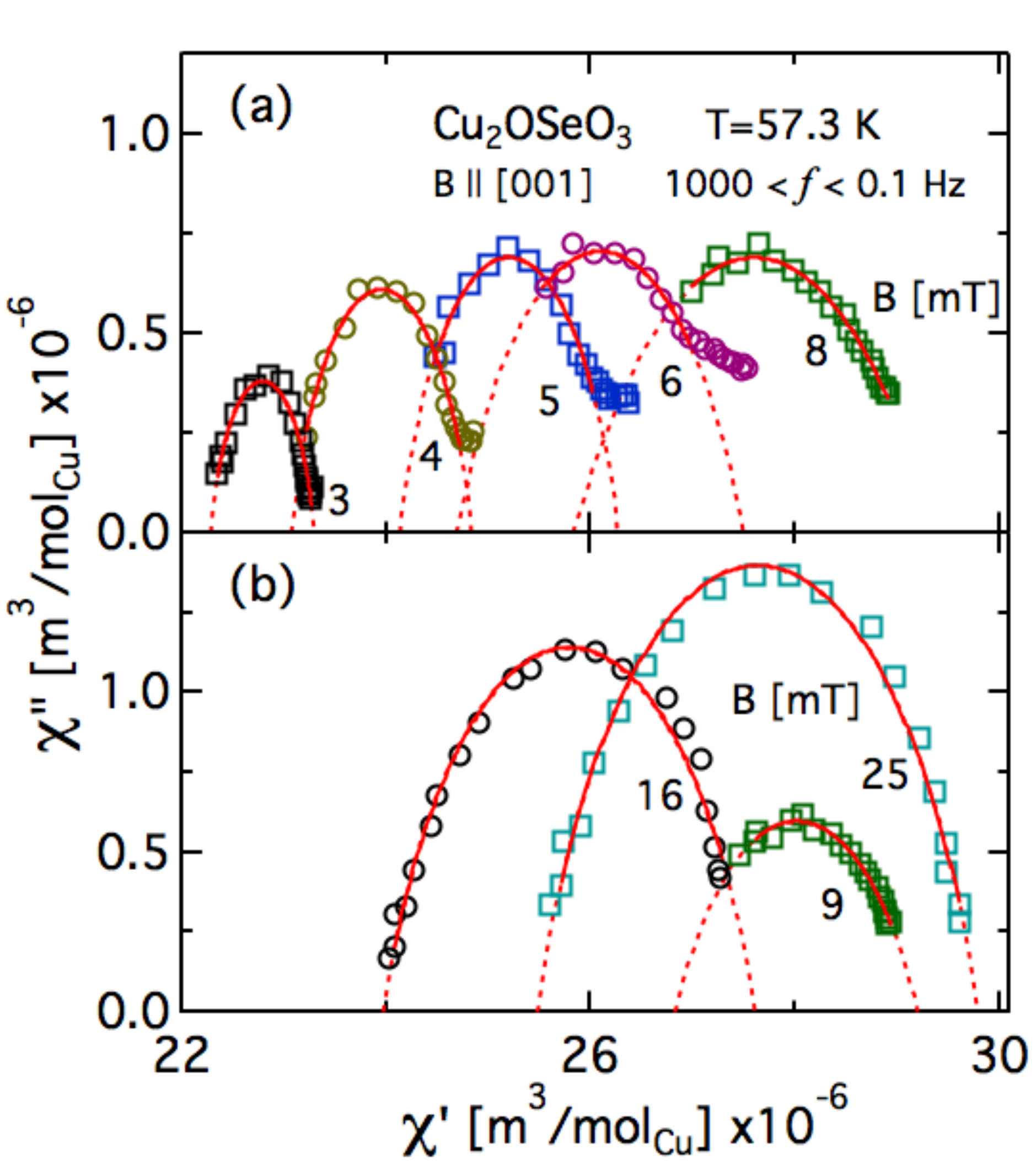}
\caption{
Cole-Cole plots of \cuse~for various magnetic fields at $T=57.3$~K. The dashed lines represent the fits of eq.~\ref{eq:colecole} to the data.}
\label{fig:ColeColePlot}
\end{center}
\end{figure}
\begin{equation}
\chi(\omega)= \chi(\infty)+\frac{\chi(0)-\chi(\infty)}{1+(i\omega \tau_0)^{1-\alpha}}
\label{eq.colecole_start}
\end{equation}
with $\chi(0)$ and $\chi(\infty)$ the isothermal and adiabatic susceptibilities, $\omega=2\pi f$ the angular frequency, $\tau_0=1/(2\pi f_0)$ the characteristic relaxation time and $\alpha$ a parameter that accounts for the width of the relaxation frequencies distribution, $\alpha=1$ for an infinitely broad distribution and $\alpha=0$ for a single relaxation process. Thus $\alpha=0$ corresponds to a simple exponential and $\alpha>0$ to a stretched exponential relaxation, which can be attributed to a distribution of energy barriers in a phase-space landscape \cite{Campbell:1986}. Eq. \ref{eq.colecole_start} can be decomposed in the in- and out-of-phase components \cite{Hueser86, Dekker} :
%
\begin{eqnarray}
\label{eq:freqdis}
\chi'(\omega)  & = &\chi(\infty) + \label{eq:Chi'Distribution} \\
 & &\frac{ A_0 \; [1+(\omega\tau_0)^{1-\alpha}\sin(\pi\alpha/2)]}{1+2(\omega\tau_0)^{1-\alpha}\sin(\pi \alpha/2 )+(\omega\tau_0)^{2(1-\alpha)} }  \notag \\
\chi''(\omega) & = & \frac{ A_0 \; \cos(\pi\alpha/2) \; (\omega\tau_0)^{1-\alpha}}{1+2(\omega\tau_0)^{1-\alpha}\sin(\pi\alpha/2)+(\omega\tau_0)^{2(1-\alpha)}   } \label{eq:Chi''Distribution}
\end{eqnarray}
with $A_0=\chi(0)-\chi(\infty)$. The fit of these equations to the data, leads to the solid lines in Fig.~\ref{fig:chiVsFreqVariousB} and the parameters, $f_0 =1/(2\pi \tau_0)$, $\alpha$ and $A_0$ are plotted as a function of the magnetic field in Fig.~\ref{fig:fitParametersT57p3K}. 

The characteristic frequency $f_0$ shown in Fig.~\ref{fig:fitParametersT57p3K}(a) is derived from the maxima of $\chi''$ varies non-monotonically with $B$. First a maximum develops slightly above $B_{C1}$, indicating the acceleration of the dynamics at the border between the helical and conical phases. Then clear minima, reflecting a slowing down of the relaxation, mark the limits of the A-phase $B_{A1}$ and $B_{A2}$ respectively. 

The parameter $\alpha$ given in Fig.~\ref{fig:fitParametersT57p3K}(b) shows no clear trends, but remains non-zero over the whole magnetic field range, as previously found\cite{Levatic14}. This implies a stretched exponential relaxation similar to spin glasses\cite{Hueser86, Dekker} and is in agreement with the glassiness found by electron microscopy for \cuse\cite{Rajeswari} and other systems with similarly long helices\cite{Milde}. 

The pre-factor $A_0=\chi(0)-\chi(\infty)$ given in Fig.~\ref{fig:fitParametersT57p3K}(c) shows clear maxima at $B_{C1}$, $B_{A1}$ and $B_{A2}$. Consequently at these characteristic fields the difference between $\chi(0)$ and $\chi(\infty)$ is the strongest and the magnetic relaxation phenomena are most prominent. 

For $B<10$~mT, which includes $B_{C1}$, the fit of $\chi'$ and $\chi''$ does not lead to the same values for $\alpha$ and $A_0$, indicating the existence of an additional process. Indeed in this magnetic field range an almost frequency independent component, designated by $\chi''_0$ and shown in Fig.~\ref{fig:fitParametersT57p3K}(d) exists that is most visible on $\chi''$ and is at the origin of the different $A_0$ values derived from $\chi'$ and $\chi''$. As it will be discussed below it is also seen on the Cole-Cole plots. 

The magnetic field dependence of $f_0$, $\alpha$, $A_0$ and $\chi''_0$ derived from $\chi''$ is given in Fig.~\ref{fig:fit_Params_vs_B.pdf} for some selected temperatures. The figure shows the same trends as Fig.~\ref{fig:fitParametersT57p3K}. The relaxation involves indeed macroscopic characteristic times, which reflect rearrangements over large magnetic volumes.
We note that all maxima and minima of $f_0$ and $A_0$ are correlated with the phase boundaries $B_{C1}$, $B_{A1}$ and $B_{A2}$.

Fig. \ref{fig:fit_Params_vs_B.pdf}(a) shows that $f_0$ is almost temperature independent at $B_{C1}$ but varies strongly with temperature at $B_{A1}$ and $B_{A2}$. At these magnetic fields a simple Arrhenius law $f_0=A~exp(-E/k_B T)$ implies a linear dependence of log($f_0$) on $1/T$. As seen in Fig.~\ref{fig:fit_Params_vs_B.pdf} (e) this is in general not the case. Furthermore, an Arrhenius fit to the data leads to un-physically large energy barriers of the order of 10$^4$~K. Thus the temperature dependence of the relaxation cannot be accounted for by a simple thermal activation picture.

\section{Cole-Cole Analysis}
\label{sec:Cole-Cole}
The Cole-Cole formalism interrelates $\chi'$ and $\chi''$ and from eq. \ref{eq:Chi'Distribution} and \ref{eq:Chi''Distribution} one deduces:
\begin{eqnarray}
\label{eq:colecole}
&\chi'' (\omega) &= -B_0 +\bigg[ B_0^2 +  A_0 \, \big(\chi'(\omega)-\chi(\infty)\big) - \\
& & - \big( \chi'(\omega)-\chi(\infty)\big)^2  \bigg]^{1/2}\notag 
\end{eqnarray}
with $B_0 =A_0 \tan (\pi \alpha /2)  /2$. This relation has three free parameters, $\chi(\infty)$, $A_0$ and $\alpha$, and corresponds to the equation of a circular arc centred at the maximum of $\chi''$, where $\omega\tau_0=1 $. Resulting Cole-Cole plots are shown in Fig.~\ref{fig:ColeColePlot}, where the ZFC data of Fig.~\ref{fig:chiVsFreqVariousB} have been replotted for selected magnetic fields. The Cole-Cole plots for one relaxation process are symmetric and centred at $\chi' = [\chi(0)+\chi(\infty)]/2$, where $\chi''$ is maximum. As the parameter $\alpha$ is proportional to the ratio of the bottom width to the height of the curves, the plots directly evidence the existence of a distribution of relaxation times. 

The data fall on a circular arc as expected by eq. \ref{eq:colecole} and are well described for all magnetic fields expect for 4~mT $\lesssim B \lesssim$ 8~mT, where significant deviations at the highest values of $\chi'$, which correspond to the lowest frequencies, are observed. This is indeed the magnetic field range, where eq. \ref{eq:Chi'Distribution} and \ref{eq:Chi''Distribution} do not give consistent results for both $\chi'$ and $\chi''$ as witnessed by the different values of $A_0$ and $\alpha$ shown in Fig.~\ref{fig:fitParametersT57p3K} and the frequency independent term $\chi''_0$ that must be considered to properly account for $\chi''$. These deviations reveal the existence of additional relaxation mechanisms, the origin of which is unknown. In analogy to ferromagnets we speculate that they may be due to domain wall motion and co-exist with the main dynamic process associated with the helical to conical phase transition. Indeed these deviations disappear above $B_{C1}$ where the system becomes mono-domain under the influence of the strong external magnetic field.
%
\begin{figure*}
\begin{center}
\includegraphics[width= .9 \textwidth]{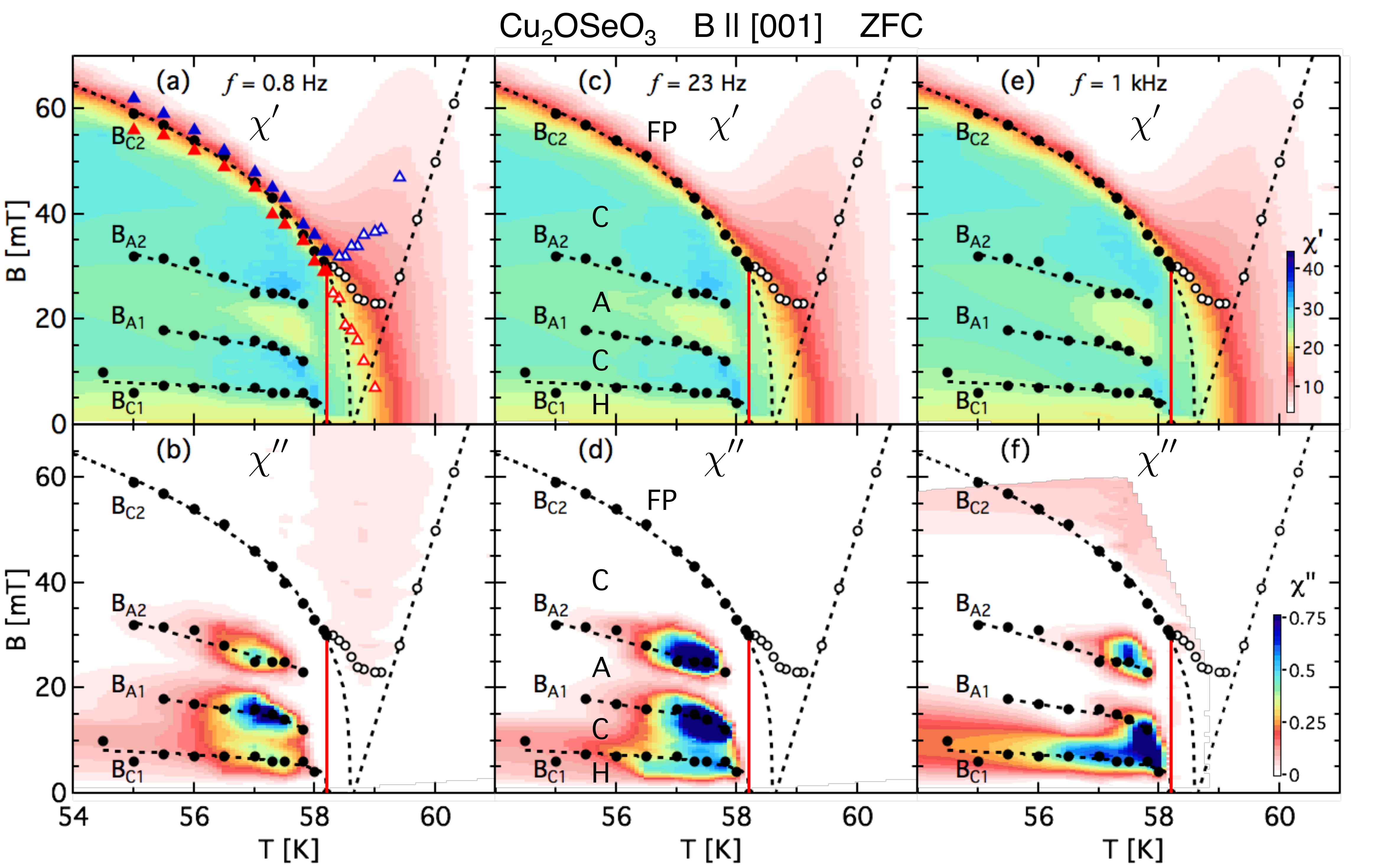}
\caption{Contour plots of ZFC $\chi'$ and $\chi''$ of \cuse~measured at the frequencies $f=0.8$~Hz (a and b) , 23~Hz (c and d) and 1~kHz (e and f) displayed versus temperature and magnetic field. The values given at the colour codes correspond to m$^3$/mol$_{\text{Cu}}\times$10$^{-6}$. The vertical red lines indicate $T_C$ at $B=0$. The characteristic magnetic fields $B_{C1}$, $B_{A1}$, $B_{A2}$ and $B_{C2}$ determined at 0.8 Hz are also indicated. The lower and upper boundaries of the transition at $B_{C2}$ are illustrated in panel (a) by the blue and red triangles determined from the inflection points of  $d\chi'/dB$. Above $T_C$ the extrema of the first and second derivatives of $\chi'$ are illustrated with open symbols to distinguish from $B_{C2}$ below $T_C$.
In the panels (c) and (d) we identify the following phases: H for helical, C for conical, A for the A-phase and FP for the field polarised one. }
\label{fig:All_Phase_diagrams}
\end{center}
\end{figure*}
\section{Phase Diagram}
\label{sec:Phase_Diagram}
The magnetic field, temperature and frequency dependence of $\chi'$ and $\chi''$ are summarised by the contour plots of Fig.~\ref{fig:All_Phase_diagrams} for $f=$ 0.8~Hz, 23~Hz and 1~kHz. The frequency has a weak effect on $\chi'$ in contrast to $\chi''$. The contour plots also illustrate the differences in the frequency dependence around $B_{C1}$ on one side and around $B_{A1}$ or $B_{A2}$ on the other side, addressed at the previous sections, e.g. see Fig.~\ref{fig:chiVsBvariousFreq}. Consequently, the strongest $\chi''$ in Fig.~\ref{fig:All_Phase_diagrams} is for $f=$ 23~Hz at $B_{A1}$ or $B_{A2}$, and for $f=$ 1~kHz at $B_{C1}$. At this highest frequency the phase boundaries seen in $\chi''$ almost obliterate.  

Very weak signals in $\chi''$ appear for $B>B_{C2}$ in the panels (b) and (f). The origin of the $\chi''$ at $f$ = 0.8~Hz is unclear and we plan to investigate this feature more in detail in the future. We note that similar features have been found also at very low frequencies in MnSi and the soliton lattice system Cr$_{1/3}$NbS$_2$\cite{Tsuruta2016}. On the other hand, the signal at 1~kHz may be associated with the spin dynamics in the field-polarised phase. In this phase the magnetic field is strong enough to overcome the DM interaction and the behaviour crosses-over to ferromagnetism with clear susceptibility maxima developing above $T_C$. These shift to higher temperatures with increasing magnetic field, as shown in Fig.~\ref{fig:MvsB_plots}(d) and in the $\chi'$ contour plots of Fig.~\ref{fig:All_Phase_diagrams}(a), (c) and (e). 

Below $T_C$ the temperature dependence of $B_{C2}$ is described by the power law $B_{C2} \propto (T-T_0)^{0.318\pm0.03}$ with $T_0=58.6 \pm 0.15$~K, i.e. $\sim T_C+0.4$~K. The extrapolation of this power law to low fields is given by the black dashed lines in Fig.~\ref{fig:All_Phase_diagrams} and is in good agreement with the contour plots of $\chi'$. 
The lower and upper magnetic field limits deduced from the inflection points of $d\chi'/dB$, as illustrated by the red dotted lines in Fig.\ref{fig:definitionBoundaries}, are also given by the blue and red triangles in Fig.~\ref{fig:All_Phase_diagrams}(a). These are very close to $B_{C2}$ deduced from the first derivative of $\chi'$ (black dots in Fig.~\ref{fig:All_Phase_diagrams}(a)) and follow the same power law. Thus the result is not affected by the specific criterion used and the sharpness of the transition.
 
Above $T_C$, the steep change of $\chi'$ at $B_{C2}$, reflected by the sharp minimum of the first derivative $d\chi'/dB$ in Fig.~\ref{fig:definitionBoundaries} broadens with increasing temperature. To distinguish from $B_{C2}$ below $T_C$, the inflection point of $\chi'$ above $T_C$ is illustrated with open circles. It goes through a minimum at $\sim 59$~K and for higher temperatures increases linearly with temperature, extrapolating to $T_0$ = 58.6 K at zero field. This linear temperature dependence of the inflection points of $\chi'$ at high fields can also be derived from a magnetisation Brillouin function. In this case, similarly to the Curie-Weiss law for the susceptibility, $T_0$ corresponds to the mean-field average of the interactions and is equivalent to a Curie-Weiss temperature. 

For $B<B_{C2}$ the plots of Fig.~\ref{fig:All_Phase_diagrams} reveal the areas associated with the helical, conical and A-phases respectively. On the $\chi'$ plots, Fig.~\ref{fig:All_Phase_diagrams}(a),(c) and (e), the yellowish area below the $B_{C1}$ line corresponds to the helical phase, whereas the light blue areas between $B_{C1}$ and $B_{A1}$ and between $B_{A2}$ and $B_{C2}$ to the conical phase. Between $B_{A1}$ and $B_{A2}$, a yellowish area with almost a triangular shape at low frequencies stands for the centre of the A-phase. The high temperature limits of the A-phase are better defined than the low temperature ones, where below $\sim$57~K a rather gradual crossover from the A-phase to the conical is observed. 

Remarkably neither the high temperature nor the low temperature limits of the A-phase are seen on $\chi''$, panels (b), (d) and (f) of Fig.~\ref{fig:All_Phase_diagrams}, in contrast to the clear delimitation of the low and upper magnetic field limits $B_{A1}$ and $B_{A2}$ respectively. We note that this result does not depend on the way the measurements were performed: temperature scans at constant magnetic fields (FC) and magnetic field scans at constant temperatures (ZFC) give exactly the same results within the experimental accuracy. Thus the temperature induced transitions into and out of the A-phase are not equivalent to the magnetic field induced ones. In MnSi the (1st order phase transition) upper temperature limits of the A-phase are clearly seen by specific heat but not in $\chi''$ for $f=$1~kHz\cite{Bauer:2013fw}. This might also explain the absence of $\chi''$ for the temperature limit of the A-phase in \cuse~ as well. 

The absence of a clear low temperature boundary of the A-phase on $\chi''$ may be another indication for a gradual crossover to the conical phase seen by $\chi'$. In such a case the magnetic relaxation phenomena might be spread over a broad temperature range and the resulting $\chi''$ signal might become too weak to be detected. 

\section{Conclusion}
\label{Conclusion}

The analysis of the DC magnetization and AC suceptibility of \cuse~ as a function of the magnetic field, instead of the temperature as in  previous studies provides a quantitative approach to the phase diagram. The investigation of the $B-T$ phase diagram of \cuse~by DC magnetisation and AC susceptibility shows that the borders between the different phases (helical, conical and A-phase) are not sharp but their exact position depends on the specific technique and criteria used.   The frequency dependence of $\chi''$, between the helical and conical phases at $B_{C1}$ and between the conical and A-Phases at $B_{A1}$ and $B_{A2}$, is governed by almost macroscopic relaxation times, which reach some milliseconds and may be attributed to rearrangements over large magnetic volumes. An additional relaxation process has also been found, which appears only at very low frequencies and around $B_{C1}$. The strongly non-exponential relaxation bears similarities with spin glasses and is in-line with the glassy behaviour reported by electron microscopy in \cuse~and other systems with similarly long helices. The dynamical phenomena discussed  at the previous sections could be at the origin of the different phase boundaries reported in the literature not only for \cuse~but also for other systems of the same family including the reference chiral magnet MnSi.

\begin{acknowledgements}
The authors would like to thank M. Mostovoy and G. R. Blake for fruitful discussions. F. Qian acknowledges financial support from China Scholarship Council. 
\end{acknowledgements}


\begin{thebibliography}{37}%
\makeatletter
\providecommand \@ifxundefined [1]{%
 \@ifx{#1\undefined}
}%
\providecommand \@ifnum [1]{%
 \ifnum #1\expandafter \@firstoftwo
 \else \expandafter \@secondoftwo
 \fi
}%
\providecommand \@ifx [1]{%
 \ifx #1\expandafter \@firstoftwo
 \else \expandafter \@secondoftwo
 \fi
}%
\providecommand \natexlab [1]{#1}%
\providecommand \enquote  [1]{``#1''}%
\providecommand \bibnamefont  [1]{#1}%
\providecommand \bibfnamefont [1]{#1}%
\providecommand \citenamefont [1]{#1}%
\providecommand \href@noop [0]{\@secondoftwo}%
\providecommand \href [0]{\begingroup \@sanitize@url \@href}%
\providecommand \@href[1]{\@@startlink{#1}\@@href}%
\providecommand \@@href[1]{\endgroup#1\@@endlink}%
\providecommand \@sanitize@url [0]{\catcode `\\12\catcode `\$12\catcode
  `\&12\catcode `\#12\catcode `\^12\catcode `\_12\catcode `\%12\relax}%
\providecommand \@@startlink[1]{}%
\providecommand \@@endlink[0]{}%
\providecommand \url  [0]{\begingroup\@sanitize@url \@url }%
\providecommand \@url [1]{\endgroup\@href {#1}{\urlprefix }}%
\providecommand \urlprefix  [0]{URL }%
\providecommand \Eprint [0]{\href }%
\providecommand \doibase [0]{http://dx.doi.org/}%
\providecommand \selectlanguage [0]{\@gobble}%
\providecommand \bibinfo  [0]{\@secondoftwo}%
\providecommand \bibfield  [0]{\@secondoftwo}%
\providecommand \translation [1]{[#1]}%
\providecommand \BibitemOpen [0]{}%
\providecommand \bibitemStop [0]{}%
\providecommand \bibitemNoStop [0]{.\EOS\space}%
\providecommand \EOS [0]{\spacefactor3000\relax}%
\providecommand \BibitemShut  [1]{\csname bibitem#1\endcsname}%
\let\auto@bib@innerbib\@empty
\bibitem [{\citenamefont {Dzyaloshinsky}(1958)}]{Dzyaloshinsky:1958vq}%
  \BibitemOpen
  \bibfield  {author} {\bibinfo {author} {\bibfnamefont {I.}~\bibnamefont
  {Dzyaloshinsky}},\ }\href@noop {} {\bibfield  {journal} {\bibinfo  {journal}
  {J. Phys. Chem. Solids}\ }\textbf {\bibinfo {volume} {4}},\ \bibinfo {pages}
  {241} (\bibinfo {year} {1958})}\BibitemShut {NoStop}%
\bibitem [{\citenamefont {Moriya}(1960)}]{Moriya:1960uf}%
  \BibitemOpen
  \bibfield  {author} {\bibinfo {author} {\bibfnamefont {T.}~\bibnamefont
  {Moriya}},\ }\href@noop {} {\bibfield  {journal} {\bibinfo  {journal}
  {Physical review}\ }\textbf {\bibinfo {volume} {120}},\ \bibinfo {pages} {91}
  (\bibinfo {year} {1960})}\BibitemShut {NoStop}%
\bibitem [{\citenamefont {Bogdanov}\ and\ \citenamefont
  {Yablonskii}(1989)}]{bogdanov1989}%
  \BibitemOpen
  \bibfield  {author} {\bibinfo {author} {\bibfnamefont {A.}~\bibnamefont
  {Bogdanov}}\ and\ \bibinfo {author} {\bibfnamefont {D.}~\bibnamefont
  {Yablonskii}},\ }\href@noop {} {\bibfield  {journal} {\bibinfo  {journal}
  {Zh. Eksp. Teor. Fiz}\ }\textbf {\bibinfo {volume} {95}},\ \bibinfo {pages}
  {182} (\bibinfo {year} {1989})}\BibitemShut {NoStop}%
\bibitem [{\citenamefont {Bogdanov}\ and\ \citenamefont
  {Hubert}(1994)}]{bogdanov1994}%
  \BibitemOpen
  \bibfield  {author} {\bibinfo {author} {\bibfnamefont {A.}~\bibnamefont
  {Bogdanov}}\ and\ \bibinfo {author} {\bibfnamefont {A.}~\bibnamefont
  {Hubert}},\ }\href@noop {} {\bibfield  {journal} {\bibinfo  {journal} {J.
  Magn. Magn. Mater.}\ }\textbf {\bibinfo {volume} {138}},\ \bibinfo {pages}
  {255} (\bibinfo {year} {1994})}\BibitemShut {NoStop}%
\bibitem [{\citenamefont {R{\"o}{\ss}ler}\ \emph {et~al.}(2011)\citenamefont
  {R{\"o}{\ss}ler}, \citenamefont {Leonov},\ and\ \citenamefont
  {Bogdanov}}]{Rossler:2011fs}%
  \BibitemOpen
  \bibfield  {author} {\bibinfo {author} {\bibfnamefont {U.~K.}\ \bibnamefont
  {R{\"o}{\ss}ler}}, \bibinfo {author} {\bibfnamefont {A.~A.}\ \bibnamefont
  {Leonov}}, \ and\ \bibinfo {author} {\bibfnamefont {A.~N.}\ \bibnamefont
  {Bogdanov}},\ }\href@noop {} {\bibfield  {journal} {\bibinfo  {journal} {J.
  Phys.: Conf. Ser.}\ }\textbf {\bibinfo {volume} {303}},\ \bibinfo {pages}
  {012105} (\bibinfo {year} {2011})}\BibitemShut {NoStop}%
\bibitem [{\citenamefont {Nagaosa}\ and\ \citenamefont
  {Tokura}(2013)}]{nagaosa2013}%
  \BibitemOpen
  \bibfield  {author} {\bibinfo {author} {\bibfnamefont {N.}~\bibnamefont
  {Nagaosa}}\ and\ \bibinfo {author} {\bibfnamefont {Y.}~\bibnamefont
  {Tokura}},\ }\href@noop {} {\bibfield  {journal} {\bibinfo  {journal} {Nat.
  Nanotechnol.}\ }\textbf {\bibinfo {volume} {8}},\ \bibinfo {pages} {899}
  (\bibinfo {year} {2013})}\BibitemShut {NoStop}%
\bibitem [{\citenamefont {M{\"u}hlbauer}\ \emph {et~al.}(2009)\citenamefont
  {M{\"u}hlbauer}, \citenamefont {Binz}, \citenamefont {Jonietz}, \citenamefont
  {Pfleiderer}, \citenamefont {Rosch}, \citenamefont {Neubauer}, \citenamefont
  {Georgii},\ and\ \citenamefont {B{\"o}ni}}]{Muhlbauer:2009bc}%
  \BibitemOpen
  \bibfield  {author} {\bibinfo {author} {\bibfnamefont {S.}~\bibnamefont
  {M{\"u}hlbauer}}, \bibinfo {author} {\bibfnamefont {B.}~\bibnamefont {Binz}},
  \bibinfo {author} {\bibfnamefont {F.}~\bibnamefont {Jonietz}}, \bibinfo
  {author} {\bibfnamefont {C.}~\bibnamefont {Pfleiderer}}, \bibinfo {author}
  {\bibfnamefont {A.}~\bibnamefont {Rosch}}, \bibinfo {author} {\bibfnamefont
  {A.}~\bibnamefont {Neubauer}}, \bibinfo {author} {\bibfnamefont
  {R.}~\bibnamefont {Georgii}}, \ and\ \bibinfo {author} {\bibfnamefont
  {P.}~\bibnamefont {B{\"o}ni}},\ }\href@noop {} {\bibfield  {journal}
  {\bibinfo  {journal} {Science}\ }\textbf {\bibinfo {volume} {323}},\ \bibinfo
  {pages} {915} (\bibinfo {year} {2009})}\BibitemShut {NoStop}%
\bibitem [{\citenamefont {Tonomura}\ \emph {et~al.}(2012)\citenamefont
  {Tonomura}, \citenamefont {Yu}, \citenamefont {Yanagisawa}, \citenamefont
  {Matsuda}, \citenamefont {Onose}, \citenamefont {Kanazawa}, \citenamefont
  {Park},\ and\ \citenamefont {Tokura}}]{Tonomura:2012ep}%
  \BibitemOpen
  \bibfield  {author} {\bibinfo {author} {\bibfnamefont {A.}~\bibnamefont
  {Tonomura}}, \bibinfo {author} {\bibfnamefont {X.}~\bibnamefont {Yu}},
  \bibinfo {author} {\bibfnamefont {K.}~\bibnamefont {Yanagisawa}}, \bibinfo
  {author} {\bibfnamefont {T.}~\bibnamefont {Matsuda}}, \bibinfo {author}
  {\bibfnamefont {Y.}~\bibnamefont {Onose}}, \bibinfo {author} {\bibfnamefont
  {N.}~\bibnamefont {Kanazawa}}, \bibinfo {author} {\bibfnamefont {H.~S.}\
  \bibnamefont {Park}}, \ and\ \bibinfo {author} {\bibfnamefont
  {Y.}~\bibnamefont {Tokura}},\ }\href@noop {} {\bibfield  {journal} {\bibinfo
  {journal} {Nano Lett.}\ }\textbf {\bibinfo {volume} {12}},\ \bibinfo {pages}
  {1673} (\bibinfo {year} {2012})}\BibitemShut {NoStop}%
\bibitem [{\citenamefont {Yu}\ \emph {et~al.}(2010{\natexlab{a}})\citenamefont
  {Yu}, \citenamefont {Kanazawa}, \citenamefont {Onose}, \citenamefont
  {Kimoto}, \citenamefont {Zhang}, \citenamefont {Ishiwata}, \citenamefont
  {Matsui},\ and\ \citenamefont {Tokura}}]{Yu:2010hr}%
  \BibitemOpen
  \bibfield  {author} {\bibinfo {author} {\bibfnamefont {X.~Z.}\ \bibnamefont
  {Yu}}, \bibinfo {author} {\bibfnamefont {N.}~\bibnamefont {Kanazawa}},
  \bibinfo {author} {\bibfnamefont {Y.}~\bibnamefont {Onose}}, \bibinfo
  {author} {\bibfnamefont {K.}~\bibnamefont {Kimoto}}, \bibinfo {author}
  {\bibfnamefont {W.~Z.}\ \bibnamefont {Zhang}}, \bibinfo {author}
  {\bibfnamefont {S.}~\bibnamefont {Ishiwata}}, \bibinfo {author}
  {\bibfnamefont {Y.}~\bibnamefont {Matsui}}, \ and\ \bibinfo {author}
  {\bibfnamefont {Y.}~\bibnamefont {Tokura}},\ }\href@noop {} {\bibfield
  {journal} {\bibinfo  {journal} {Nat. Mater.}\ }\textbf {\bibinfo {volume}
  {10}},\ \bibinfo {pages} {106} (\bibinfo {year}
  {2010}{\natexlab{a}})}\BibitemShut {NoStop}%
\bibitem [{\citenamefont {Wilhelm}\ \emph {et~al.}(2011)\citenamefont
  {Wilhelm}, \citenamefont {Baenitz}, \citenamefont {Schmidt}, \citenamefont
  {R{\"o}{\ss}ler}, \citenamefont {Leonov},\ and\ \citenamefont
  {Bogdanov}}]{Wilhelm:2011jva}%
  \BibitemOpen
  \bibfield  {author} {\bibinfo {author} {\bibfnamefont {H.}~\bibnamefont
  {Wilhelm}}, \bibinfo {author} {\bibfnamefont {M.}~\bibnamefont {Baenitz}},
  \bibinfo {author} {\bibfnamefont {M.}~\bibnamefont {Schmidt}}, \bibinfo
  {author} {\bibfnamefont {U.~K.}\ \bibnamefont {R{\"o}{\ss}ler}}, \bibinfo
  {author} {\bibfnamefont {A.~A.}\ \bibnamefont {Leonov}}, \ and\ \bibinfo
  {author} {\bibfnamefont {A.~N.}\ \bibnamefont {Bogdanov}},\ }\href@noop {}
  {\bibfield  {journal} {\bibinfo  {journal} {Phys. Rev. Lett.}\ }\textbf
  {\bibinfo {volume} {107}},\ \bibinfo {pages} {127203} (\bibinfo {year}
  {2011})}\BibitemShut {NoStop}%
\bibitem [{\citenamefont {Moskvin}\ \emph {et~al.}(2013)\citenamefont
  {Moskvin}, \citenamefont {Grigoriev}, \citenamefont {Dyadkin}, \citenamefont
  {Eckerlebe}, \citenamefont {Baenitz}, \citenamefont {Schmidt},\ and\
  \citenamefont {Wilhelm}}]{Moskvin:2013kf}%
  \BibitemOpen
  \bibfield  {author} {\bibinfo {author} {\bibfnamefont {E.}~\bibnamefont
  {Moskvin}}, \bibinfo {author} {\bibfnamefont {S.}~\bibnamefont {Grigoriev}},
  \bibinfo {author} {\bibfnamefont {V.}~\bibnamefont {Dyadkin}}, \bibinfo
  {author} {\bibfnamefont {H.}~\bibnamefont {Eckerlebe}}, \bibinfo {author}
  {\bibfnamefont {M.}~\bibnamefont {Baenitz}}, \bibinfo {author} {\bibfnamefont
  {M.}~\bibnamefont {Schmidt}}, \ and\ \bibinfo {author} {\bibfnamefont
  {H.}~\bibnamefont {Wilhelm}},\ }\href@noop {} {\bibfield  {journal} {\bibinfo
   {journal} {Phys. Rev. Lett.}\ }\textbf {\bibinfo {volume} {110}},\ \bibinfo
  {pages} {077207} (\bibinfo {year} {2013})}\BibitemShut {NoStop}%
\bibitem [{\citenamefont {Yu}\ \emph {et~al.}(2010{\natexlab{b}})\citenamefont
  {Yu}, \citenamefont {Onose}, \citenamefont {Kanazawa}, \citenamefont {Park},
  \citenamefont {Han}, \citenamefont {Matsui}, \citenamefont {Nagaosa},\ and\
  \citenamefont {Tokura}}]{Yu:2010iu}%
  \BibitemOpen
  \bibfield  {author} {\bibinfo {author} {\bibfnamefont {X.~Z.}\ \bibnamefont
  {Yu}}, \bibinfo {author} {\bibfnamefont {Y.}~\bibnamefont {Onose}}, \bibinfo
  {author} {\bibfnamefont {N.}~\bibnamefont {Kanazawa}}, \bibinfo {author}
  {\bibfnamefont {J.~H.}\ \bibnamefont {Park}}, \bibinfo {author}
  {\bibfnamefont {J.~H.}\ \bibnamefont {Han}}, \bibinfo {author} {\bibfnamefont
  {Y.}~\bibnamefont {Matsui}}, \bibinfo {author} {\bibfnamefont
  {N.}~\bibnamefont {Nagaosa}}, \ and\ \bibinfo {author} {\bibfnamefont
  {Y.}~\bibnamefont {Tokura}},\ }\href@noop {} {\bibfield  {journal} {\bibinfo
  {journal} {Nature}\ }\textbf {\bibinfo {volume} {465}},\ \bibinfo {pages}
  {901} (\bibinfo {year} {2010}{\natexlab{b}})}\BibitemShut {NoStop}%
\bibitem [{\citenamefont {M{\"u}nzer}\ \emph {et~al.}(2010)\citenamefont
  {M{\"u}nzer}, \citenamefont {Neubauer}, \citenamefont {Adams}, \citenamefont
  {M{\"u}hlbauer}, \citenamefont {Franz}, \citenamefont {Jonietz},
  \citenamefont {Georgii}, \citenamefont {B{\"o}ni}, \citenamefont {Pedersen},
  \citenamefont {Schmidt}, \citenamefont {Rosch},\ and\ \citenamefont
  {Pfleiderer}}]{munzer2010}%
  \BibitemOpen
  \bibfield  {author} {\bibinfo {author} {\bibfnamefont {W.}~\bibnamefont
  {M{\"u}nzer}}, \bibinfo {author} {\bibfnamefont {A.}~\bibnamefont
  {Neubauer}}, \bibinfo {author} {\bibfnamefont {T.}~\bibnamefont {Adams}},
  \bibinfo {author} {\bibfnamefont {S.}~\bibnamefont {M{\"u}hlbauer}}, \bibinfo
  {author} {\bibfnamefont {C.}~\bibnamefont {Franz}}, \bibinfo {author}
  {\bibfnamefont {F.}~\bibnamefont {Jonietz}}, \bibinfo {author} {\bibfnamefont
  {R.}~\bibnamefont {Georgii}}, \bibinfo {author} {\bibfnamefont
  {P.}~\bibnamefont {B{\"o}ni}}, \bibinfo {author} {\bibfnamefont
  {B.}~\bibnamefont {Pedersen}}, \bibinfo {author} {\bibfnamefont
  {M.}~\bibnamefont {Schmidt}}, \bibinfo {author} {\bibfnamefont
  {A.}~\bibnamefont {Rosch}}, \ and\ \bibinfo {author} {\bibfnamefont
  {C.}~\bibnamefont {Pfleiderer}},\ }\href@noop {} {\bibfield  {journal}
  {\bibinfo  {journal} {Phys. Rev. B}\ }\textbf {\bibinfo {volume} {81}},\
  \bibinfo {pages} {041203(R)} (\bibinfo {year} {2010})}\BibitemShut {NoStop}%
\bibitem [{\citenamefont {Park}\ \emph {et~al.}(2014)\citenamefont {Park},
  \citenamefont {Yu}, \citenamefont {Aizawa}, \citenamefont {Tanigaki},
  \citenamefont {Akashi}, \citenamefont {Takahashi}, \citenamefont {Matsuda},
  \citenamefont {Kanazawa}, \citenamefont {Onose}, \citenamefont {Shindo},
  \citenamefont {Tonomura},\ and\ \citenamefont {Tokura}}]{Park:2014jc}%
  \BibitemOpen
  \bibfield  {author} {\bibinfo {author} {\bibfnamefont {H.~S.}\ \bibnamefont
  {Park}}, \bibinfo {author} {\bibfnamefont {X.}~\bibnamefont {Yu}}, \bibinfo
  {author} {\bibfnamefont {S.}~\bibnamefont {Aizawa}}, \bibinfo {author}
  {\bibfnamefont {T.}~\bibnamefont {Tanigaki}}, \bibinfo {author}
  {\bibfnamefont {T.}~\bibnamefont {Akashi}}, \bibinfo {author} {\bibfnamefont
  {Y.}~\bibnamefont {Takahashi}}, \bibinfo {author} {\bibfnamefont
  {T.}~\bibnamefont {Matsuda}}, \bibinfo {author} {\bibfnamefont
  {N.}~\bibnamefont {Kanazawa}}, \bibinfo {author} {\bibfnamefont
  {Y.}~\bibnamefont {Onose}}, \bibinfo {author} {\bibfnamefont
  {D.}~\bibnamefont {Shindo}}, \bibinfo {author} {\bibfnamefont
  {A.}~\bibnamefont {Tonomura}}, \ and\ \bibinfo {author} {\bibfnamefont
  {Y.}~\bibnamefont {Tokura}},\ }\href@noop {} {\bibfield  {journal} {\bibinfo
  {journal} {Nat. Nanotechnol.}\ }\textbf {\bibinfo {volume} {9}},\ \bibinfo
  {pages} {337} (\bibinfo {year} {2014})}\BibitemShut {NoStop}%
\bibitem [{\citenamefont {Seki}\ \emph
  {et~al.}(2012{\natexlab{a}})\citenamefont {Seki}, \citenamefont {Yu},
  \citenamefont {Ishiwata},\ and\ \citenamefont {Tokura}}]{Seki:2012ie}%
  \BibitemOpen
  \bibfield  {author} {\bibinfo {author} {\bibfnamefont {S.}~\bibnamefont
  {Seki}}, \bibinfo {author} {\bibfnamefont {X.~Z.}\ \bibnamefont {Yu}},
  \bibinfo {author} {\bibfnamefont {S.}~\bibnamefont {Ishiwata}}, \ and\
  \bibinfo {author} {\bibfnamefont {Y.}~\bibnamefont {Tokura}},\ }\href@noop {}
  {\bibfield  {journal} {\bibinfo  {journal} {Science}\ }\textbf {\bibinfo
  {volume} {336}},\ \bibinfo {pages} {198} (\bibinfo {year}
  {2012}{\natexlab{a}})}\BibitemShut {NoStop}%
\bibitem [{\citenamefont {Seki}\ \emph
  {et~al.}(2012{\natexlab{b}})\citenamefont {Seki}, \citenamefont {Kim},
  \citenamefont {Inosov}, \citenamefont {Georgii}, \citenamefont {Keimer},
  \citenamefont {Ishiwata},\ and\ \citenamefont {Tokura}}]{Seki:2012ch}%
  \BibitemOpen
  \bibfield  {author} {\bibinfo {author} {\bibfnamefont {S.}~\bibnamefont
  {Seki}}, \bibinfo {author} {\bibfnamefont {J.~H.}\ \bibnamefont {Kim}},
  \bibinfo {author} {\bibfnamefont {D.~S.}\ \bibnamefont {Inosov}}, \bibinfo
  {author} {\bibfnamefont {R.}~\bibnamefont {Georgii}}, \bibinfo {author}
  {\bibfnamefont {B.}~\bibnamefont {Keimer}}, \bibinfo {author} {\bibfnamefont
  {S.}~\bibnamefont {Ishiwata}}, \ and\ \bibinfo {author} {\bibfnamefont
  {Y.}~\bibnamefont {Tokura}},\ }\href@noop {} {\bibfield  {journal} {\bibinfo
  {journal} {Phys. Rev. B}\ }\textbf {\bibinfo {volume} {85}},\ \bibinfo
  {pages} {220406(R)} (\bibinfo {year} {2012}{\natexlab{b}})}\BibitemShut
  {NoStop}%
\bibitem [{\citenamefont {Adams}\ \emph {et~al.}(2012)\citenamefont {Adams},
  \citenamefont {Chacon}, \citenamefont {Wagner}, \citenamefont {Bauer},
  \citenamefont {Brandl}, \citenamefont {Pedersen}, \citenamefont {Berger},
  \citenamefont {Lemmens},\ and\ \citenamefont {Pfleiderer}}]{Adams12}%
  \BibitemOpen
  \bibfield  {author} {\bibinfo {author} {\bibfnamefont {T.}~\bibnamefont
  {Adams}}, \bibinfo {author} {\bibfnamefont {A.}~\bibnamefont {Chacon}},
  \bibinfo {author} {\bibfnamefont {M.}~\bibnamefont {Wagner}}, \bibinfo
  {author} {\bibfnamefont {A.}~\bibnamefont {Bauer}}, \bibinfo {author}
  {\bibfnamefont {G.}~\bibnamefont {Brandl}}, \bibinfo {author} {\bibfnamefont
  {B.}~\bibnamefont {Pedersen}}, \bibinfo {author} {\bibfnamefont
  {H.}~\bibnamefont {Berger}}, \bibinfo {author} {\bibfnamefont
  {P.}~\bibnamefont {Lemmens}}, \ and\ \bibinfo {author} {\bibfnamefont
  {C.}~\bibnamefont {Pfleiderer}},\ }\href@noop {} {\bibfield  {journal}
  {\bibinfo  {journal} {Phys. Rev. Lett.}\ }\textbf {\bibinfo {volume} {108}},\
  \bibinfo {pages} {237204} (\bibinfo {year} {2012})}\BibitemShut {NoStop}%
\bibitem [{\citenamefont {White}\ \emph {et~al.}(2012)\citenamefont {White},
  \citenamefont {Levati{\'c}}, \citenamefont {Omrani}, \citenamefont
  {Egetenmeyer}, \citenamefont {Pr{\v s}a}, \citenamefont {{\v Z}ivkovi{\'c}},
  \citenamefont {Gavilano}, \citenamefont {Kohlbrecher}, \citenamefont
  {Bartkowiak}, \citenamefont {Berger},\ and\ \citenamefont
  {R{\o}nnow}}]{White12}%
  \BibitemOpen
  \bibfield  {author} {\bibinfo {author} {\bibfnamefont {J.~S.}\ \bibnamefont
  {White}}, \bibinfo {author} {\bibfnamefont {I.}~\bibnamefont {Levati{\'c}}},
  \bibinfo {author} {\bibfnamefont {A.~A.}\ \bibnamefont {Omrani}}, \bibinfo
  {author} {\bibfnamefont {N.}~\bibnamefont {Egetenmeyer}}, \bibinfo {author}
  {\bibfnamefont {K.}~\bibnamefont {Pr{\v s}a}}, \bibinfo {author}
  {\bibfnamefont {I.}~\bibnamefont {{\v Z}ivkovi{\'c}}}, \bibinfo {author}
  {\bibfnamefont {J.~L.}\ \bibnamefont {Gavilano}}, \bibinfo {author}
  {\bibfnamefont {J.}~\bibnamefont {Kohlbrecher}}, \bibinfo {author}
  {\bibfnamefont {M.}~\bibnamefont {Bartkowiak}}, \bibinfo {author}
  {\bibfnamefont {H.}~\bibnamefont {Berger}}, \ and\ \bibinfo {author}
  {\bibfnamefont {H.~M.}\ \bibnamefont {R{\o}nnow}},\ }\href@noop {} {\bibfield
   {journal} {\bibinfo  {journal} {J. Phys.: Condens. Matter}\ }\textbf
  {\bibinfo {volume} {24}},\ \bibinfo {pages} {432201} (\bibinfo {year}
  {2012})}\BibitemShut {NoStop}%
\bibitem [{\citenamefont {White}\ \emph {et~al.}(2014)\citenamefont {White},
  \citenamefont {Pr{\v s}a}, \citenamefont {Huang}, \citenamefont {Omrani},
  \citenamefont {{\v Z}ivkovi{\'c}}, \citenamefont {Bartkowiak}, \citenamefont
  {Berger}, \citenamefont {Magrez}, \citenamefont {Gavilano}, \citenamefont
  {Nagy}, \citenamefont {Zang},\ and\ \citenamefont {R{\o}nnow}}]{White14}%
  \BibitemOpen
  \bibfield  {author} {\bibinfo {author} {\bibfnamefont {J.~S.}\ \bibnamefont
  {White}}, \bibinfo {author} {\bibfnamefont {K.}~\bibnamefont {Pr{\v s}a}},
  \bibinfo {author} {\bibfnamefont {P.}~\bibnamefont {Huang}}, \bibinfo
  {author} {\bibfnamefont {A.~A.}\ \bibnamefont {Omrani}}, \bibinfo {author}
  {\bibfnamefont {I.}~\bibnamefont {{\v Z}ivkovi{\'c}}}, \bibinfo {author}
  {\bibfnamefont {M.}~\bibnamefont {Bartkowiak}}, \bibinfo {author}
  {\bibfnamefont {H.}~\bibnamefont {Berger}}, \bibinfo {author} {\bibfnamefont
  {A.}~\bibnamefont {Magrez}}, \bibinfo {author} {\bibfnamefont {J.~L.}\
  \bibnamefont {Gavilano}}, \bibinfo {author} {\bibfnamefont {G.}~\bibnamefont
  {Nagy}}, \bibinfo {author} {\bibfnamefont {J.}~\bibnamefont {Zang}}, \ and\
  \bibinfo {author} {\bibfnamefont {H.~M.}\ \bibnamefont {R{\o}nnow}},\
  }\href@noop {} {\bibfield  {journal} {\bibinfo  {journal} {Phys. Rev. Lett.}\
  }\textbf {\bibinfo {volume} {113}},\ \bibinfo {pages} {107203} (\bibinfo
  {year} {2014})}\BibitemShut {NoStop}%
\bibitem [{\citenamefont {Bos}\ \emph {et~al.}(2008)\citenamefont {Bos},
  \citenamefont {Colin},\ and\ \citenamefont {Palstra}}]{Bos08}%
  \BibitemOpen
  \bibfield  {author} {\bibinfo {author} {\bibfnamefont {J.-W.~G.}\
  \bibnamefont {Bos}}, \bibinfo {author} {\bibfnamefont {C.~V.}\ \bibnamefont
  {Colin}}, \ and\ \bibinfo {author} {\bibfnamefont {T.~T.~M.}\ \bibnamefont
  {Palstra}},\ }\href {\doibase 10.1103/PhysRevB.78.094416} {\bibfield
  {journal} {\bibinfo  {journal} {Phys. Rev. B}\ }\textbf {\bibinfo {volume}
  {78}},\ \bibinfo {pages} {094416} (\bibinfo {year} {2008})}\BibitemShut
  {NoStop}%
\bibitem [{\citenamefont {Belesi}\ \emph {et~al.}(2010)\citenamefont {Belesi},
  \citenamefont {Rousochatzakis}, \citenamefont {Wu},\ and\ \citenamefont
  {Berger}}]{Belesi:2010vz}%
  \BibitemOpen
  \bibfield  {author} {\bibinfo {author} {\bibfnamefont {M.}~\bibnamefont
  {Belesi}}, \bibinfo {author} {\bibfnamefont {I.}~\bibnamefont
  {Rousochatzakis}}, \bibinfo {author} {\bibfnamefont {H.~C.}\ \bibnamefont
  {Wu}}, \ and\ \bibinfo {author} {\bibfnamefont {H.}~\bibnamefont {Berger}},\
  }\href@noop {} {\bibfield  {journal} {\bibinfo  {journal} {Phys. Rev. B}\ }
  (\bibinfo {year} {2010})}\BibitemShut {NoStop}%
\bibitem [{\citenamefont {Levati{\'c}}\ \emph {et~al.}(2014)\citenamefont
  {Levati{\'c}}, \citenamefont {{\v S}urija}, \citenamefont {Berger},\ and\
  \citenamefont {{\v Z}ivkovi{\'c}}}]{Levatic14}%
  \BibitemOpen
  \bibfield  {author} {\bibinfo {author} {\bibfnamefont {I.}~\bibnamefont
  {Levati{\'c}}}, \bibinfo {author} {\bibfnamefont {V.}~\bibnamefont {{\v
  S}urija}}, \bibinfo {author} {\bibfnamefont {H.}~\bibnamefont {Berger}}, \
  and\ \bibinfo {author} {\bibfnamefont {I.}~\bibnamefont {{\v
  Z}ivkovi{\'c}}},\ }\href@noop {} {\bibfield  {journal} {\bibinfo  {journal}
  {Phys. Rev. B}\ }\textbf {\bibinfo {volume} {90}},\ \bibinfo {pages} {224412}
  (\bibinfo {year} {2014})}\BibitemShut {NoStop}%
\bibitem [{\citenamefont {Miller}\ \emph {et~al.}(2010)\citenamefont {Miller},
  \citenamefont {Xu}, \citenamefont {Berger}, \citenamefont {Knowles},
  \citenamefont {Arenas}, \citenamefont {Meisel},\ and\ \citenamefont
  {Tanner}}]{Miller:2010fu}%
  \BibitemOpen
  \bibfield  {author} {\bibinfo {author} {\bibfnamefont {K.~H.}\ \bibnamefont
  {Miller}}, \bibinfo {author} {\bibfnamefont {X.~S.}\ \bibnamefont {Xu}},
  \bibinfo {author} {\bibfnamefont {H.}~\bibnamefont {Berger}}, \bibinfo
  {author} {\bibfnamefont {E.~S.}\ \bibnamefont {Knowles}}, \bibinfo {author}
  {\bibfnamefont {D.~J.}\ \bibnamefont {Arenas}}, \bibinfo {author}
  {\bibfnamefont {M.~W.}\ \bibnamefont {Meisel}}, \ and\ \bibinfo {author}
  {\bibfnamefont {D.~B.}\ \bibnamefont {Tanner}},\ }\href@noop {} {\bibfield
  {journal} {\bibinfo  {journal} {Phys. Rev. B}\ }\textbf {\bibinfo {volume}
  {82}},\ \bibinfo {pages} {144107} (\bibinfo {year} {2010})}\BibitemShut
  {NoStop}%
\bibitem [{\citenamefont {{\v Z}ivkovi{\'c}}\ \emph {et~al.}(2012)\citenamefont
  {{\v Z}ivkovi{\'c}}, \citenamefont {Paji{\'c}}, \citenamefont {Ivek},\ and\
  \citenamefont {Berger}}]{Zivkovi12}%
  \BibitemOpen
  \bibfield  {author} {\bibinfo {author} {\bibfnamefont {I.}~\bibnamefont {{\v
  Z}ivkovi{\'c}}}, \bibinfo {author} {\bibfnamefont {D.}~\bibnamefont
  {Paji{\'c}}}, \bibinfo {author} {\bibfnamefont {T.}~\bibnamefont {Ivek}}, \
  and\ \bibinfo {author} {\bibfnamefont {H.}~\bibnamefont {Berger}},\
  }\href@noop {} {\bibfield  {journal} {\bibinfo  {journal} {Phys. Rev. B}\
  }\textbf {\bibinfo {volume} {85}},\ \bibinfo {pages} {224402} (\bibinfo
  {year} {2012})}\BibitemShut {NoStop}%
\bibitem [{\citenamefont {Stishov}\ \emph {et~al.}(2007)\citenamefont
  {Stishov}, \citenamefont {Petrova}, \citenamefont {Khasanov}, \citenamefont
  {Panova}, \citenamefont {Shikov}, \citenamefont {Lashley}, \citenamefont
  {Wu},\ and\ \citenamefont {Lograsso}}]{2007PhRvB..76e2405S}%
  \BibitemOpen
  \bibfield  {author} {\bibinfo {author} {\bibfnamefont {S.~M.}\ \bibnamefont
  {Stishov}}, \bibinfo {author} {\bibfnamefont {A.~E.}\ \bibnamefont
  {Petrova}}, \bibinfo {author} {\bibfnamefont {S.}~\bibnamefont {Khasanov}},
  \bibinfo {author} {\bibfnamefont {G.~K.}\ \bibnamefont {Panova}}, \bibinfo
  {author} {\bibfnamefont {A.~A.}\ \bibnamefont {Shikov}}, \bibinfo {author}
  {\bibfnamefont {J.~C.}\ \bibnamefont {Lashley}}, \bibinfo {author}
  {\bibfnamefont {D.}~\bibnamefont {Wu}}, \ and\ \bibinfo {author}
  {\bibfnamefont {T.~A.}\ \bibnamefont {Lograsso}},\ }\href@noop {} {\bibfield
  {journal} {\bibinfo  {journal} {Physical Review B}\ }\textbf {\bibinfo
  {volume} {76}},\ \bibinfo {pages} {052405} (\bibinfo {year}
  {2007})}\BibitemShut {NoStop}%
\bibitem [{\citenamefont {Stishov}\ and\ \citenamefont
  {Petrova}(2016)}]{2016arXiv160606922S}%
  \BibitemOpen
  \bibfield  {author} {\bibinfo {author} {\bibfnamefont {S.~M.}\ \bibnamefont
  {Stishov}}\ and\ \bibinfo {author} {\bibfnamefont {A.~E.}\ \bibnamefont
  {Petrova}},\ }\href@noop {} {\bibfield  {journal} {\bibinfo  {journal}
  {arXiv.org}\ ,\ \bibinfo {pages} {arXiv:1606.06922}} (\bibinfo {year}
  {2016})},\ \Eprint {http://arxiv.org/abs/1606.06922} {1606.06922}
  \BibitemShut {NoStop}%
\bibitem [{\citenamefont {{\v Z}ivkovi{\'c}}\ \emph {et~al.}(2014)\citenamefont
  {{\v Z}ivkovi{\'c}}, \citenamefont {White}, \citenamefont {R{\o}nnow},
  \citenamefont {Pr{\v s}a},\ and\ \citenamefont {Berger}}]{Zivkovi14}%
  \BibitemOpen
  \bibfield  {author} {\bibinfo {author} {\bibfnamefont {I.}~\bibnamefont {{\v
  Z}ivkovi{\'c}}}, \bibinfo {author} {\bibfnamefont {J.~S.}\ \bibnamefont
  {White}}, \bibinfo {author} {\bibfnamefont {H.~M.}\ \bibnamefont
  {R{\o}nnow}}, \bibinfo {author} {\bibfnamefont {K.}~\bibnamefont {Pr{\v
  s}a}}, \ and\ \bibinfo {author} {\bibfnamefont {H.}~\bibnamefont {Berger}},\
  }\href@noop {} {\bibfield  {journal} {\bibinfo  {journal} {Phys. Rev. B}\
  }\textbf {\bibinfo {volume} {89}},\ \bibinfo {pages} {060401(R)} (\bibinfo
  {year} {2014})}\BibitemShut {NoStop}%
\bibitem [{\citenamefont {Seki}\ \emph
  {et~al.}(2012{\natexlab{c}})\citenamefont {Seki}, \citenamefont {Ishiwata},\
  and\ \citenamefont {Tokura}}]{Seki12}%
  \BibitemOpen
  \bibfield  {author} {\bibinfo {author} {\bibfnamefont {S.}~\bibnamefont
  {Seki}}, \bibinfo {author} {\bibfnamefont {S.}~\bibnamefont {Ishiwata}}, \
  and\ \bibinfo {author} {\bibfnamefont {Y.}~\bibnamefont {Tokura}},\
  }\href@noop {} {\bibfield  {journal} {\bibinfo  {journal} {Phys. Rev. B}\
  }\textbf {\bibinfo {volume} {86}},\ \bibinfo {pages} {060403(R)} (\bibinfo
  {year} {2012}{\natexlab{c}})}\BibitemShut {NoStop}%
\bibitem [{\citenamefont {Bauer}\ and\ \citenamefont
  {Pfleiderer}(2012)}]{Bauer:2012cw}%
  \BibitemOpen
  \bibfield  {author} {\bibinfo {author} {\bibfnamefont {A.}~\bibnamefont
  {Bauer}}\ and\ \bibinfo {author} {\bibfnamefont {C.}~\bibnamefont
  {Pfleiderer}},\ }\href@noop {} {\bibfield  {journal} {\bibinfo  {journal}
  {Phys. Rev. B}\ }\textbf {\bibinfo {volume} {85}},\ \bibinfo {pages} {214418}
  (\bibinfo {year} {2012})}\BibitemShut {NoStop}%
\bibitem [{\citenamefont {Cole}\ and\ \citenamefont {Cole}(1941)}]{cole_cole}%
  \BibitemOpen
  \bibfield  {author} {\bibinfo {author} {\bibfnamefont {K.~S.}\ \bibnamefont
  {Cole}}\ and\ \bibinfo {author} {\bibfnamefont {R.~H.}\ \bibnamefont
  {Cole}},\ }\href@noop {} {\bibfield  {journal} {\bibinfo  {journal} {J. Chem.
  Phys.}\ }\textbf {\bibinfo {volume} {9}},\ \bibinfo {pages} {341} (\bibinfo
  {year} {1941})}\BibitemShut {NoStop}%
\bibitem [{\citenamefont {Huser}\ \emph {et~al.}(1986)\citenamefont {Huser},
  \citenamefont {van Duyneveldt}, \citenamefont {Nieuwenhuys},\ and\
  \citenamefont {Mydosh}}]{Hueser86}%
  \BibitemOpen
  \bibfield  {author} {\bibinfo {author} {\bibfnamefont {D.}~\bibnamefont
  {Huser}}, \bibinfo {author} {\bibfnamefont {A.~J.}\ \bibnamefont {van
  Duyneveldt}}, \bibinfo {author} {\bibfnamefont {G.~J.}\ \bibnamefont
  {Nieuwenhuys}}, \ and\ \bibinfo {author} {\bibfnamefont {J.~A.}\ \bibnamefont
  {Mydosh}},\ }\href@noop {} {\bibfield  {journal} {\bibinfo  {journal} {J.
  Phys. C: Solid State Phys.}\ }\textbf {\bibinfo {volume} {19}},\ \bibinfo
  {pages} {3697} (\bibinfo {year} {1986})}\BibitemShut {NoStop}%
\bibitem [{\citenamefont {Campbell}(1986)}]{Campbell:1986}%
  \BibitemOpen
  \bibfield  {author} {\bibinfo {author} {\bibfnamefont {I.~A.}\ \bibnamefont
  {Campbell}},\ }\href@noop {} {\bibfield  {journal} {\bibinfo  {journal}
  {Phys. Rev. B}\ }\textbf {\bibinfo {volume} {33}},\ \bibinfo {pages} {3587}
  (\bibinfo {year} {1986})}\BibitemShut {NoStop}%
\bibitem [{\citenamefont {Dekker}\ \emph {et~al.}(1989)\citenamefont {Dekker},
  \citenamefont {Arts}, \citenamefont {Dewijn}, \citenamefont {Vanduyneveldt},\
  and\ \citenamefont {Mydosh}}]{Dekker}%
  \BibitemOpen
  \bibfield  {author} {\bibinfo {author} {\bibfnamefont {C.}~\bibnamefont
  {Dekker}}, \bibinfo {author} {\bibfnamefont {A.}~\bibnamefont {Arts}},
  \bibinfo {author} {\bibfnamefont {H.~W.}\ \bibnamefont {Dewijn}}, \bibinfo
  {author} {\bibfnamefont {A.~J.}\ \bibnamefont {Vanduyneveldt}}, \ and\
  \bibinfo {author} {\bibfnamefont {J.~A.}\ \bibnamefont {Mydosh}},\
  }\href@noop {} {\bibfield  {journal} {\bibinfo  {journal} {Phys. Rev. B}\
  }\textbf {\bibinfo {volume} {40}},\ \bibinfo {pages} {11243} (\bibinfo {year}
  {1989})}\BibitemShut {NoStop}%
\bibitem [{\citenamefont {Rajeswari}\ \emph {et~al.}(2015)\citenamefont
  {Rajeswari}, \citenamefont {Huang}, \citenamefont {Mancini}, \citenamefont
  {Murooka}, \citenamefont {Latychevskaia}, \citenamefont {McGrouther},
  \citenamefont {Cantoni}, \citenamefont {Baldini}, \citenamefont {White},
  \citenamefont {Magrez}, \citenamefont {Giamarchi}, \citenamefont
  {R{\o}nnow},\ and\ \citenamefont {Carbone}}]{Rajeswari}%
  \BibitemOpen
  \bibfield  {author} {\bibinfo {author} {\bibfnamefont {J.}~\bibnamefont
  {Rajeswari}}, \bibinfo {author} {\bibfnamefont {P.}~\bibnamefont {Huang}},
  \bibinfo {author} {\bibfnamefont {G.~F.}\ \bibnamefont {Mancini}}, \bibinfo
  {author} {\bibfnamefont {Y.}~\bibnamefont {Murooka}}, \bibinfo {author}
  {\bibfnamefont {T.}~\bibnamefont {Latychevskaia}}, \bibinfo {author}
  {\bibfnamefont {D.}~\bibnamefont {McGrouther}}, \bibinfo {author}
  {\bibfnamefont {M.}~\bibnamefont {Cantoni}}, \bibinfo {author} {\bibfnamefont
  {E.}~\bibnamefont {Baldini}}, \bibinfo {author} {\bibfnamefont {J.~S.}\
  \bibnamefont {White}}, \bibinfo {author} {\bibfnamefont {A.}~\bibnamefont
  {Magrez}}, \bibinfo {author} {\bibfnamefont {T.}~\bibnamefont {Giamarchi}},
  \bibinfo {author} {\bibfnamefont {H.~M.}\ \bibnamefont {R{\o}nnow}}, \ and\
  \bibinfo {author} {\bibfnamefont {F.}~\bibnamefont {Carbone}},\ }\href@noop
  {} {\bibfield  {journal} {\bibinfo  {journal} {P Natl Acad Sci USA}\ }\textbf
  {\bibinfo {volume} {112}},\ \bibinfo {pages} {14212} (\bibinfo {year}
  {2015})}\BibitemShut {NoStop}%
\bibitem [{\citenamefont {Milde}\ \emph {et~al.}(2013)\citenamefont {Milde},
  \citenamefont {Kohler}, \citenamefont {Seidel}, \citenamefont {Eng},
  \citenamefont {Bauer}, \citenamefont {Chacon}, \citenamefont {Kindervater},
  \citenamefont {Muhlbauer}, \citenamefont {Pfleiderer}, \citenamefont
  {Buhrandt}, \citenamefont {Schutte},\ and\ \citenamefont {Rosch}}]{Milde}%
  \BibitemOpen
  \bibfield  {author} {\bibinfo {author} {\bibfnamefont {P.}~\bibnamefont
  {Milde}}, \bibinfo {author} {\bibfnamefont {D.}~\bibnamefont {Kohler}},
  \bibinfo {author} {\bibfnamefont {J.}~\bibnamefont {Seidel}}, \bibinfo
  {author} {\bibfnamefont {L.~M.}\ \bibnamefont {Eng}}, \bibinfo {author}
  {\bibfnamefont {A.}~\bibnamefont {Bauer}}, \bibinfo {author} {\bibfnamefont
  {A.}~\bibnamefont {Chacon}}, \bibinfo {author} {\bibfnamefont
  {J.}~\bibnamefont {Kindervater}}, \bibinfo {author} {\bibfnamefont
  {S.}~\bibnamefont {Muhlbauer}}, \bibinfo {author} {\bibfnamefont
  {C.}~\bibnamefont {Pfleiderer}}, \bibinfo {author} {\bibfnamefont
  {S.}~\bibnamefont {Buhrandt}}, \bibinfo {author} {\bibfnamefont
  {C.}~\bibnamefont {Schutte}}, \ and\ \bibinfo {author} {\bibfnamefont
  {A.}~\bibnamefont {Rosch}},\ }\href@noop {} {\bibfield  {journal} {\bibinfo
  {journal} {Science}\ }\textbf {\bibinfo {volume} {340}},\ \bibinfo {pages}
  {1076} (\bibinfo {year} {2013})}\BibitemShut {NoStop}%
\bibitem [{\citenamefont {Tsuruta}\ \emph {et~al.}(2016)\citenamefont
  {Tsuruta}, \citenamefont {Mito}, \citenamefont {Kousaka}, \citenamefont
  {Akimitsu}, \citenamefont {Kishine},\ and\ \citenamefont
  {Inoue}}]{Tsuruta2016}%
  \BibitemOpen
  \bibfield  {author} {\bibinfo {author} {\bibfnamefont {K.}~\bibnamefont
  {Tsuruta}}, \bibinfo {author} {\bibfnamefont {M.}~\bibnamefont {Mito}},
  \bibinfo {author} {\bibfnamefont {Y.}~\bibnamefont {Kousaka}}, \bibinfo
  {author} {\bibfnamefont {J.}~\bibnamefont {Akimitsu}}, \bibinfo {author}
  {\bibfnamefont {J.}~\bibnamefont {Kishine}}, \ and\ \bibinfo {author}
  {\bibfnamefont {K.}~\bibnamefont {Inoue}},\ }\href@noop {} {\enquote
  {\bibinfo {title} {{Nonlinear magnetic responses in skyrmion phase of MnSi
  and chiral-soliton-lattice phase of Cr$_{1/3}$NbS$_2$}},}\ }\bibinfo
  {howpublished} {poster presentation at $\chi$-Mag2016, Hiroshima, Japan}
  (\bibinfo {year} {2016})\BibitemShut {NoStop}%
\bibitem [{\citenamefont {Bauer}\ \emph {et~al.}(2013)\citenamefont {Bauer},
  \citenamefont {Garst},\ and\ \citenamefont {Pfleiderer}}]{Bauer:2013fw}%
  \BibitemOpen
  \bibfield  {author} {\bibinfo {author} {\bibfnamefont {A.}~\bibnamefont
  {Bauer}}, \bibinfo {author} {\bibfnamefont {M.}~\bibnamefont {Garst}}, \ and\
  \bibinfo {author} {\bibfnamefont {C.}~\bibnamefont {Pfleiderer}},\
  }\href@noop {} {\bibfield  {journal} {\bibinfo  {journal} {Phys. Rev. Lett.}\
  }\textbf {\bibinfo {volume} {110}},\ \bibinfo {pages} {177207} (\bibinfo
  {year} {2013})}\BibitemShut {NoStop}%
\end{thebibliography}%
%

\end{document}